\documentclass[aps,twocolumn,prb,amsmath,amssymb]{revtex4-1}

\usepackage{graphicx}
\usepackage{color}
\usepackage{natbib}
\usepackage{hyperref}
\usepackage{notes2bib}
\usepackage{mathbbol}
\usepackage{ulem}

\def\dg{{\;\dagger}}
\def\m{\mathcal}
\def\doubleunderline#1{\underline{\underline{#1}}}

\begin{document}

\parindent=0pt

\title{Level anticrossing effect in single-level or multilevel double quantum dots:\\
Electrical conductance, zero-frequency charge susceptibility and Seebeck coefficient}
\author{M. Lavagna$^{1}$}
\author{V. Talbo$^{1}$}
\author{T.Q. Duong$^{2,3}$}
\author{A. Cr\'epieux$^{2}$}
\affiliation{$^1$ Univ. Grenoble Alpes, CEA, IRIG, PHELIQS, F-38000 Grenoble, France}
\affiliation{$^2$ Aix Marseille Univ, Universit\'e de Toulon, CNRS, CPT, Marseille, France}
\affiliation{$^3$ Univ. Lille, CNRS, Centrale Lille, Yncr\'ea ISEN, Univ. Polytechnique Hauts-de-France, UMR 8520, IEMN, F-59000 Lille, France
}

\begin{abstract}

{ We study electrical and thermoelectrical properties for a double quantum dot system. We consider the cases of both single-level and multilevel quantum dots whatever the way they are coupled, either in a series or in a parallel arrangement. The calculations are performed by using the nonequilibrium Green function theory.} In the case of a single-level double quantum dot, the problem is exactly solvable whereas for a multilevel double quantum dot, an analytical solution is obtained in the limit of energy-independent hopping integrals. { We present a detailed discussion about} the dependences of electrical conductance, zero-frequency charge susceptibility and Seebeck coefficient on the gate voltages applied to the dots, allowing us to derive the charge stability diagram. The findings are in agreement with the experimental observations notably with the occurrence of successive sign changes of the Seebeck coefficient when varying the gate voltages. We interpret the results in terms of the bonding and antibonding states produced by the level anticrossing effect which occurs in the presence of a finite interdot coupling. We show that at equilibrium the boundary lines between the domains with different dot occupancies in the charge stability diagram, take place when the bonding and antibonding state levels are aligned with the chemical potentials in the leads. Finally the total dot occupancy is found to be considerably reduced in the case in parallel compared with the case in series, { whenever} the level energies in each dot are equal. We interpret this dip as a direct manifestation of the interference effects occurring in the presence of the two electronic transmission paths provided by each dot. 

\end{abstract}

\maketitle



\section{Introduction}

The study of double quantum dot (DQD) has been the focus of an increasing number of research works in the last years both theoretically and experimentally. One of the main reasons explaining these developments are that DQDs are promising candidates to build {  quantum bits of spin, i.e.} spin qubits\cite{Loss1998,Huang2019}, which surpass charge qubits because of their longer coherence time\cite{Kawakami2014}. Moreover, a DQD is a readily tunable system\cite{VanDerWiel,Hanson2007,Zwanenburg2013}: the number of electrons in each of the two dots can be controlled by varying gate voltages located at proximity. Proposals to probe and drive the spin and charge states in DQDs have also been made\cite{Barthel2010,Parafilo2018,Maslova2019}. Note that the second dot is used under certain circumstances to control the spin state in the first dot benefiting from the Pauli spin blockade effect\cite{Ono2002}. Initially, experimental DQDs were built from GaAs heterostructures\cite{VanDerVaart1995,Oosterkamp1998,Elzerman2003,Petta2004,Petta2005,Pioro-Ladriere2005,Koppens2006} but one has witnessed in the last five years the development of studies in Si-based DQDs which have the advantage over the former one to present a longer spin coherence time\cite{Villis2014,House2015,Voisin2016,Maurand2016,Corna2017,Crippa2017a,Crippa2018,Crippa2018a,Ibberson2018a,Perron2017,Mi2018,Hensen2020}. The serial-coupling of the two dots is by far the geometry that has been the most studied. However, the case of two parallel coupled dots is interesting too since it may give rise to interference effects\cite{Mourokh2004,Meden2006,Li2019} or other specific effects\cite{Guevara2003,Lu2005,Kashcheyevs2007,Urban2009,Wang2011a,BulnesCuetara2016,Zhang2017,Protsenko2017}. In this paper, both geometries of serial-coupled and parallel-coupled DQDs are considered.

From the theoretical side, the electrical transport properties in DQDs have been widely studied {  by using various methods going from scattering matrix theory\cite{Rotter2004,Rotter2005,Potz2008} for noninteracting case to Master or Bloch equation approaches\cite{Stoof1996,Matveev1996,Ziegler,Hazelzet2001,Golovach2004,Schultz2009,Schaller2009} and nonequilibrium Green function methods\cite{Bulka2004,Gong2006,Sztenkiel2007,Levy2013,Kagan2018a,Dey2019} for more general cases. A classical theory has also been developed along which the DQD is modeled as a network of resistors and capacitors which mimic the tunnel and electrostatic couplings between dots and leads\cite{VanDerWiel,Hanson2007,Zwanenburg2013}. The obtained results for the overall evolution of the linear electrical conductance as a function of gate voltages are as follows: (i) at weak interdot coupling, conductance peaks are observed at the nodes of a square lattice in the  $(\varepsilon_1,\varepsilon_2)$ phase space, where $\varepsilon_1$ and $\varepsilon_2$ are the level energies of each of the two dots, delimiting domains with different integer occupancies in the charge stability diagram, (ii) at intermediate interdot coupling, the nodes separate into pairs of triple points, corresponding to the deformation of the square lattice into a honeycomb lattice in the  $(\varepsilon_1,\varepsilon_2)$ phase space, and (iii) at strong interdot coupling, the triple point separation reaches its maximum and the DQD behaves as a single dot with an occupancy  $\langle\widehat N_1\rangle+\langle\widehat N_2\rangle$, where $\langle\widehat N_1\rangle$ and $\langle\widehat N_2\rangle$ are the average rate of electronic occupancy in the dots 1 and 2 respectively. 

Charge susceptibility and thermoelectrical properties in DQDs have aroused much less attention than electrical conductance while the growing-up activity in both gate-reflectometry experiments\cite{Crippa2017a,Crippa2018a,Ahmed2018} and thermopower measurements\cite{Thierschmann2013,Thierschmann2015a,Thierschmann2015,Thierschmann2016,Thierschmann2016a} in spin qubits provides a strong motivation to intensify the theoretical efforts in that direction. From the theoretical side, one essentially refers to the works of Refs.~\onlinecite{Cottet2011,Mizuta2017,Talbo2018} where the charge susceptibility of a DQD is discussed in the context of mesoscopic admittance in either the noninteracting case or the interacting case where it is affected by the Pauli spin blockade effect. These works are in line with the seminal works of B\"uttiker and collaborators\cite{Buttiker1993,Pretre1996a} which demonstrate the importance of taking the charge susceptibility contribution into account in the mesoscopic capacitance of the system, in addition to the standard geometrical capacitances of the dots. The charge susceptibility reflects the ability of the system when it is open, to adjust the average rate of electronic occupancy in the dots to the external excitation brought by the gate voltage $V_g$. It is the relevant physical quantity behind reflectometry experiments carried out in spin qubit systems and plays a key role in the context of manipulation, coupling and readout of qubits. Theoretical works devoted to the thermoelectrical properties of DQDs have highlighted specific features such as the increase of the figure of merit when the interdot coupling is reduced\cite{BagheriTagani2012}, the decrease of the efficiency at maximum power in the presence of Coulomb interactions in the dots\cite{Eckern2019}. Some other works have focused on the effects of electron-hole symmetry\cite{Zhang2007,Tang2018} and quantum interferences\cite{Trocha2012,Karwacki2016} on thermopower. Moreover, it has been shown that a DQD constitutes a minimal thermoelectric generator\cite{Donsa2014} and the issue related to energy harvesting is a central one\cite{Juergens2013,Dare2017,Mazal2019}. All these results demonstrate the need to develop further theoretical studies on charge susceptibility and thermopower in DQDs. 

In this paper, we  simultaneously discuss electrical transport, charge susceptibility and thermoelectrical properties of a DQD connected to two reservoirs (leads) of electrons, whether the two dots are coupled in series or in parallel, and contain a single energy level or multiple energy levels. The calculations are performed by using the nonequilibrium Green function technique. We study the evolution of the electrical conductance, Seebeck coefficient, and zero-frequency charge susceptibility with the gate voltages applied to the two dots as well as the stability phase diagram, in various regimes going from weak to strong interdot coupling. The results are valid at any temperature, lead-dot and interdot couplings, bias and gate voltages. They are in qualitative agreement with experiments, with notably the succession of sign changes of the Seebeck coefficient when varying the gate voltages applied to the dots. 


The plan of the paper is the following. In Sec.~\ref{section_hamiltonian}, we present the hamiltonian describing the DQD system. We adopt a unified presentation which enables to describe the case when the dots are coupled in series as well as the case in parallel. In Sec.~\ref{section_current}, we give the analytical expressions for the electrical current which allows us to derive the linear electrical conductance and the Seebeck coefficient. In Sec.~\ref{section_charge_susceptibility}, we derive the expression for the zero-frequency charge susceptibility related to the occupancies of the dots. The numerical results obtained for conductance, Seebeck coefficient, occupancies of the dots and zero-frequency charge susceptibility are presented in Sec.~\ref{section_DQD_series} in the case when the dots are coupled in series, and in Sec.~\ref{section_DQD_parallel} when the dots are coupled in parallel. Section~\ref{section_conclusion} is a conclusion. Details about the determination of the Green functions in the dots are reported in Appendix~\ref{appendix_electrical_current}, whereas Appendix~\ref{appendix_Hdot} presents the diagonalization of the hamiltonian describing the DQD.}


\section{Model and expressions for electrical current and charge susceptibility}\label{section_model}

\subsection{Hamiltonian}\label{section_hamiltonian}

\begin{figure}[t]
\begin{center}
\includegraphics[scale=.55]{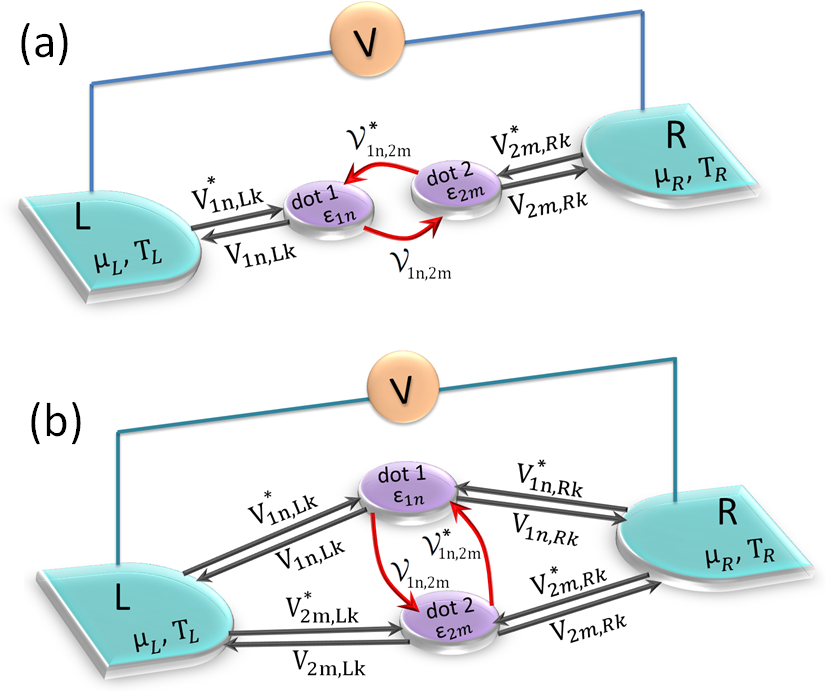}
\end{center}
\caption{
Schematic representation of the DQD coupled to $L$ and $R$ leads in the serial (a) and parallel (b) geometries.}
\label{fig1}
\end{figure} 

We consider two quantum dots 1 and 2 coupled through a tunnel barrier and connected to two metallic leads as depicted in Fig.~\ref{fig1}. Each of the two quantum dot contains $N_\varepsilon$ and $M_\varepsilon$ discrete levels of energies $\varepsilon_{1n}$ and $\varepsilon_{2m}$ respectively with $n\in[0,N_\varepsilon-1]$ and $m\in[0,M_\varepsilon-1]$. The values of $\varepsilon_{1n}$ and $\varepsilon_{2m}$  can be tuned  by varying the nearby gate voltages. The two metallic left (L) and right (R) leads are characterized by their chemical potentials $\mu_L$, $\mu_R$ and temperatures $T_L$, $T_R$ respectively. In the noninteracting case that we consider,  the hamiltonian of this DQD system writes: $\m{\widehat H} =  \m{\widehat H}_\text{dots} +\m{\widehat H}_\text{leads}  + \m{\widehat H}_\text{hop}$, where $\m{\widehat H}_\text{dots}$ and $\m{\widehat H}_\text{leads}$ are the hamiltonian of the disconnected DQD and that of the disconnected leads respectively, and $\m{\widehat H}_\text{hop}$ is the hopping hamiltonian between the dots and the leads
\begin{eqnarray}
\label{Hdot}
&&\m{\widehat H}_\text{dots} = \sum_{\substack{i=1,2\\n \in i}}\varepsilon_{in }\widehat d_{in}^\dg \widehat d_{in}
+\sum_{\substack{n\in 1\\m \in 2}}\m{V}_{1n,2m}\widehat d_{2m}^\dg \widehat d_{1n} +h.c.\\
\label{Hres}
&&\m{\widehat H}_\text{leads} = \sum_{\substack{\alpha =L,R\\k \in \alpha}} \varepsilon_{\alpha k}\widehat c_{\alpha k}^\dg \widehat c_{\alpha k}\\
\label{HT2}
&&\m{\widehat H}_\text{hop}=\sum_{\substack{\alpha =L,R\\k \in \alpha}}\sum_{\substack{i=1,2\\n \in i}}V_{in,\alpha k}\widehat c_{\alpha k}^\dg \widehat d_{in} +h.c.
\end{eqnarray}
in which $\widehat d_{in}^\dg$ and $\widehat d_{in}$ are the creation and annihilation operators of one electron in the dot $i$ with energy $\varepsilon_{in}$, the index $i$ taking the values 1 and 2. $\widehat c_{\alpha k}^\dg$ and $\widehat c_{\alpha k}$ are the creation and annihilation operators of one electron in the lead $\alpha$ with momentum $k$ and energy $\varepsilon_{\alpha k}$, the index $\alpha$ taking the values $L$ and $R$. $\m{V}_{1n,2m}$ and $V_{in,\alpha k}$ are the hopping matrix elements between the states $|1n\rangle$ and $|2m\rangle$ in the dots and those between the states $|in\rangle$ in the dot $i$ and $|\alpha k\rangle$ in the lead $\alpha$. We have $\m{V}^*_{1n,2m}=\m{V}_{2m,1n}$ and $V^*_{in,\alpha k}=V_{\alpha k,in}$. The abbreviation $h.c.$ stands for hermitian conjugate.

We want to emphasize the very general feature of the expression we introduce for the hamiltonian. It allows one to describe all the possible assemblies of two dots in a unified way: the serial assembly corresponding to the case where $V_{2m,Lk}=V_{1n,Rk}=0$ for any index $n,m$ or $k$ (depicted in Fig.~\ref{fig1}(a)), and the parallel one corresponding to the case where $V_{in,\alpha k}\ne 0$ for any index $i,n,\alpha$ or $k$  (depicted in Fig.~\ref{fig1}(b)). The spin degree of freedom can be added without any difficulty, which is essential if one wants to describe spin qubits or magnetic leads for instance. However, it will not be included in this paper since we restrict our study to a noninteracting DQD system connected to nonmagnetic leads.


\subsection{Electrical current}\label{section_current}

{ 
In this section we give the expression for the electrical current assuming that the noninteracting DQD is in a steady state. For that we start from the current operator from the lead $\alpha$ defined as $\widehat I_\alpha(t)=-e d\widehat N_\alpha(t)/dt$ where $\widehat N_\alpha(t)=\sum_{k\in\alpha}\widehat c^\dg_{\alpha k}(t)\widehat c_{\alpha k}(t)$ is the average number of electrons in the lead $\alpha$. In the steady state the derivative with respect to the time variable is given by $d\widehat N_\alpha(t)/dt=[\widehat N_\alpha(t),\m{\widehat  H}]/i\hbar$. In the limit of wide flat band and energy-independent hopping integrals, one obtains 
\begin{eqnarray}\label{I_final_maintext}
  I_\alpha=\frac{e}{h}\int_{-\infty}^{\infty}
&&\text{Tr}\Big[\doubleunderline{\Gamma}_{\;\alpha}\; \doubleunderline{\bf G}^{r}(\varepsilon)\;\doubleunderline{\Gamma}_{\;\overline{\alpha}} \;\doubleunderline{\bf G}^{a}(\varepsilon)\Big]\nonumber\\
&&\times\big(f_\alpha(\varepsilon)-f_{\overline{\alpha}}(\varepsilon)\big)d\varepsilon
\end{eqnarray}
where $\overline{\alpha}=R$ for $\alpha=L$ and $\overline{\alpha}=L$ for $\alpha=R$, $f_\alpha (\varepsilon)$ is the Fermi-Dirac distribution function in the lead $\alpha$, $ \doubleunderline{\bf G}^{r(a)}(\varepsilon)$ is the retarded (advanced) Green function matrix in the dots defined as
\begin{eqnarray}
  \doubleunderline{\bf G}^{r(a)}(\varepsilon)=\left(
\begin{array}{cc}
{\bf G}^{r(a)}_{11}(\varepsilon)&{\bf G}^{r(a)}_{12}(\varepsilon)\\
{\bf G}^{r(a)}_{21}(\varepsilon)&{\bf G}^{r(a)}_{22}(\varepsilon)
\end{array}
\right)
\end{eqnarray}
and $\doubleunderline{\Gamma}_{\;\alpha} $ is the dot-lead coupling matrix defined as
\begin{eqnarray}
\doubleunderline{\Gamma}_{\;\alpha} =2\pi\rho_\alpha\left(
\begin{array}{cc}
|V_{1\alpha}|^2 & V^*_{1\alpha}V_{2\alpha}\\
V_{1\alpha}V^*_{2\alpha} & |V_{2\alpha}|^2
\end{array}
\right)
\end{eqnarray}
with $\rho_\alpha$, the density of states in the lead $\alpha$. $\text{Tr}[\;]$ is the trace of the matrix. Note that the matrix $\doubleunderline{\Gamma}_{\;\alpha}$ is diagonal in the case of a DQD coupled in series and nondiagonal in the case in parallel. Equation~(\ref{I_final_maintext}) corresponds to the Landauer formula for the electrical current with a transmission coefficient equal to
\begin{eqnarray}\label{T_expression}
 \m{T}_{\alpha\overline{\alpha}}(\varepsilon)=\text{Tr}\Big[\doubleunderline{\Gamma}_{\;\alpha} \doubleunderline{\bf G}^{r}(\varepsilon)\doubleunderline{\Gamma}_{\;\overline{\alpha}}\doubleunderline{\bf G}^{a}(\varepsilon)\Big]
\end{eqnarray}
Appendix~\ref{appendix_electrical_current} gives the details of these calculations performed in the framework of the nonequilibrium Green function technique\cite{Caroli1971,Rammer1986,Jauho1994}.  The result for the expression of the retarded Green function matrix in the dots is
\begin{eqnarray}\label{FG_maintext}
\doubleunderline{\bf G}^{r}(\varepsilon) =
\frac{1}{D^{r}(\varepsilon)}\left(
\begin{array}{cc}
\widetilde{\bf g}_1^{r}(\varepsilon) & \widetilde{\bf g}_1^{r}(\varepsilon)\mathbb{\Sigma}_{12}^{r}(\varepsilon)\widetilde{\bf g}_2^{r}(\varepsilon)\\
\widetilde{\bf g}_2^{r}(\varepsilon)\mathbb{\Sigma}_{21}^{r}(\varepsilon)\widetilde{\bf g}_1^{r}(\varepsilon) & \widetilde{\bf g}_2^{r}(\varepsilon)
\end{array}
\right)\nonumber\\
\end{eqnarray}
where $D^{r}(\varepsilon) = 1-\widetilde{\bf g}_{1}^{r}(\varepsilon)\mathbb{\Sigma}_{12}^{r}(\varepsilon)\widetilde{\bf g}_{2}^{r}(\varepsilon)\mathbb{\Sigma}_{21}^{r}(\varepsilon)$,  ${\widetilde{\bf g}_{i}^{r}}(\varepsilon) ={\bf g}_{i}^{r} (\varepsilon)/(1 - {{{\mathbb{\Sigma}}}_{\text{hop},ii}^{r}}(\varepsilon){\bf g}_{i}^{r}(\varepsilon))$,  ${\bf g}_i^{r}(\varepsilon)$ is the retarded Green function in the disconnected dot $i$ defined as ${\bf g}_i^{r}(\varepsilon)=\sum_{n\in i}g_{in}^{r}(\varepsilon)$ with $g_{in}^{r}(\varepsilon) =  (\varepsilon - \varepsilon_{in} + i0^+)^{-1}$, and $\mathbb{\Sigma}_{ij}^{r}(\varepsilon)$ is the self-energy given by  $\mathbb{\Sigma}_{ij}^{r}(\varepsilon)= {{\mathbb{\Sigma}}}_{\text{hop},ij}^r(\varepsilon)+\delta_{j\overline{i}}\m{V}^*_{i\overline{i}}$ with ${\mathbb{\Sigma}}_{\text{hop},ij}^{r}(\varepsilon)=\sum_{\alpha=L,R}\sum_{k\in \alpha}V_{i\alpha}^*g^r_{\alpha k}(\varepsilon)V_{j\alpha}$. In the multilevel case, it has been assumed that the hopping integrals between the dots, $\mathcal{V}_{in,jm}$, and between the dots and the leads, $V_{in,αk}$, do not depend on the indices $n$, $m$ and on the state $k$.

}


\subsection{Zero-frequency charge susceptibility and dot occupancies}\label{section_charge_susceptibility}

The experimental works carried out on DQDs often focused on establishing the charge stability diagram which gives information on the charge occupancy $\langle\widehat N_i\rangle=\sum_{n\in i}\langle \widehat d_{in}^\dg \widehat d_{in}\rangle$ of each of the two dots\cite{Petta2004,Hu2007,Viennot2016,Perron2017}. One of the relevant physical quantity to discuss the charge stability diagram is the charge susceptibility which is the linear response in charge $\widehat Q(t)$ to the external excitation brought by a time-dependent gate voltage $\Delta V_g(t)$ applied to the system.  $\widehat Q(t)$, the charge accumulated on the capacitor plates ensuring the coupling between  $\Delta V_g(t)$ and the dots, is given by\cite{Cottet2011}
\begin{eqnarray}\label{def_Q}
 \widehat Q(t)=(C_1^0+C_2^0)\Delta V_g(t) \mathbb{\widehat I}-\sum_{i=1,2}\alpha_i e\Delta\widehat N_i(t)
\end{eqnarray}
where $C_1^0$ and $C_2^0$ are the { geometrical} capacitances of the totally disconnected quantum dots (closed system with $\mathcal{V}_{12}=0$ and $V_{i\alpha}=0$, $\forall i,\alpha$) and $\mathbb{\widehat I}$ is the identity operator. The last term in Eq.~(\ref{def_Q}) comes from the additional electrons in the dot $i$, denoted by $\Delta\widehat N_i(t)$, induced by  $\Delta V_g(t)$ when the dots are connected (open system), weighted by the lever-arm coefficient, $\alpha_i$ measuring the asymmetry of the capacitive coupling of the voltage generator to the dot $i$. The external excitation $\Delta V_g(t)$ introduces the additional source term $ \Delta\widehat H(t)= \widehat Q(t)\Delta V_g(t)$ in the hamiltonian of Eqs.~(\ref{Hdot})-(\ref{HT2}). From the linear response theory\cite{Fetter1971}, the expectation value $\langle\widehat Q(t)\rangle$, up to the first order in $\Delta V_g(t)$, is given by
\begin{eqnarray}\label{av_Q}
 \langle\widehat Q(t)\rangle&=&(C_1^0+C_2^0)\Delta V_g(t)\nonumber\\
&&-e^2\int_{-\infty}^{\infty}\chi_c(t,t')\Delta V_g(t')dt'
\end{eqnarray}
where $\chi_c(t,t')$ is the dynamical charge susceptibility given by the Kubo formula: $\chi_c(t,t')=\sum_{i,j=1,2}\alpha_i\alpha_j\chi_{ij}(t,t')$, with
\begin{eqnarray}
 \chi_{ij}(t,t')=i\Theta(t-t')\langle[\Delta \widehat N_{i}(t),\Delta \widehat N_{j}(t')]\rangle
\end{eqnarray}
By taking the Fourier transform of Eq.~(\ref{av_Q}), one gets $\langle\widehat Q(\omega)\rangle=C(\omega)\Delta V_g(\omega)$ where $C(\omega)=C_1^0+C_2^0-e^2 \chi_c(\omega)$ is the mesoscopic capacitance\cite{Buttiker1993,Pretre1996a} of the DQD system, with $\chi_c(\omega)=\sum_{i,j=1,2}\alpha_i\alpha_j\chi_{ij}(\omega)$. Thus, in addition to $C_1^0+C_2^0$, $C(\omega)$ contains an additional contribution, equal to $-e^2 \chi_c(\omega)$, related to the dynamical charge susceptibility defined as
\begin{eqnarray}\label{dcs}
 e^2 \chi_c(\omega)=\lim_{\Delta V_g(\omega)\rightarrow 0}\sum_{i=1,2}\alpha_i e\frac{d\langle \Delta \widehat N_i(\omega)\rangle}{d\Delta V_g(\omega)}
\end{eqnarray}
In the following we will focus on the static charge susceptibility $\chi_c(\omega=0)$. To get it, we will not make use of Eq.~(\ref{dcs}) but rather of the alternative and more direct expression given by
\begin{eqnarray}\label{exp_chi_derivative_new}
 \chi_c(\omega=0)=- \sum_{i,j=1,2}\alpha_i \alpha_j\frac{\partial \langle \widehat N_i\rangle }{\partial \varepsilon_j}
\end{eqnarray}
where $\langle \widehat N_i\rangle$ is the number of electrons in the dot $i$ at $\Delta V_g(\omega)=0$ given by
\begin{eqnarray}\label{def_N}
 \langle\widehat N_i\rangle=-\frac{i}{2\pi}\sum_{n\in i}\int_{-\infty}^{\infty} d\varepsilon G^<_{in,in}(\varepsilon)
\end{eqnarray}
Equation~(\ref{exp_chi_derivative_new}) can be readily obtained from Eq.~(\ref{dcs}) by noticing that the role of $\Delta V_g(\omega=0)$ comes down to shift the level energy of the dot $j$ according to $\varepsilon_j\rightarrow \widetilde \varepsilon_j=\varepsilon_j-\alpha_j e\Delta V_g(\omega=0)$, therefore
\begin{eqnarray}\label{eq_lim}
 &&\lim_{\Delta V_g(\omega=0)\rightarrow 0}\frac{d\langle \Delta \widehat  N_i(\omega)\rangle}{d\Delta V_g(\omega=0)}=\sum_{j=1,2}\frac{\partial\langle \Delta \widehat N_i(\omega=0)\rangle}{\partial \widetilde\varepsilon_j}\nonumber\\
&&\times\frac{d\widetilde\varepsilon_j}{d\Delta V_g(\omega=0)}=-e\sum_{j=1,2}\alpha_j\frac{\partial \widehat N_i}{\partial\varepsilon_j}
\end{eqnarray}
Incorporating Eq.~(\ref{eq_lim}) into Eq.~(\ref{dcs}), one obtains Eq.~(\ref{exp_chi_derivative_new}). We set $\alpha_1=\alpha_2=1$ in the rest of the paper. However, the influence of asymmetric capacitive couplings can be readily studied by using the results we obtain for arbitrary values of $\alpha_1$ and $\alpha_2$.

\begin{figure}[t]
\begin{center}
\includegraphics[scale=0.4]{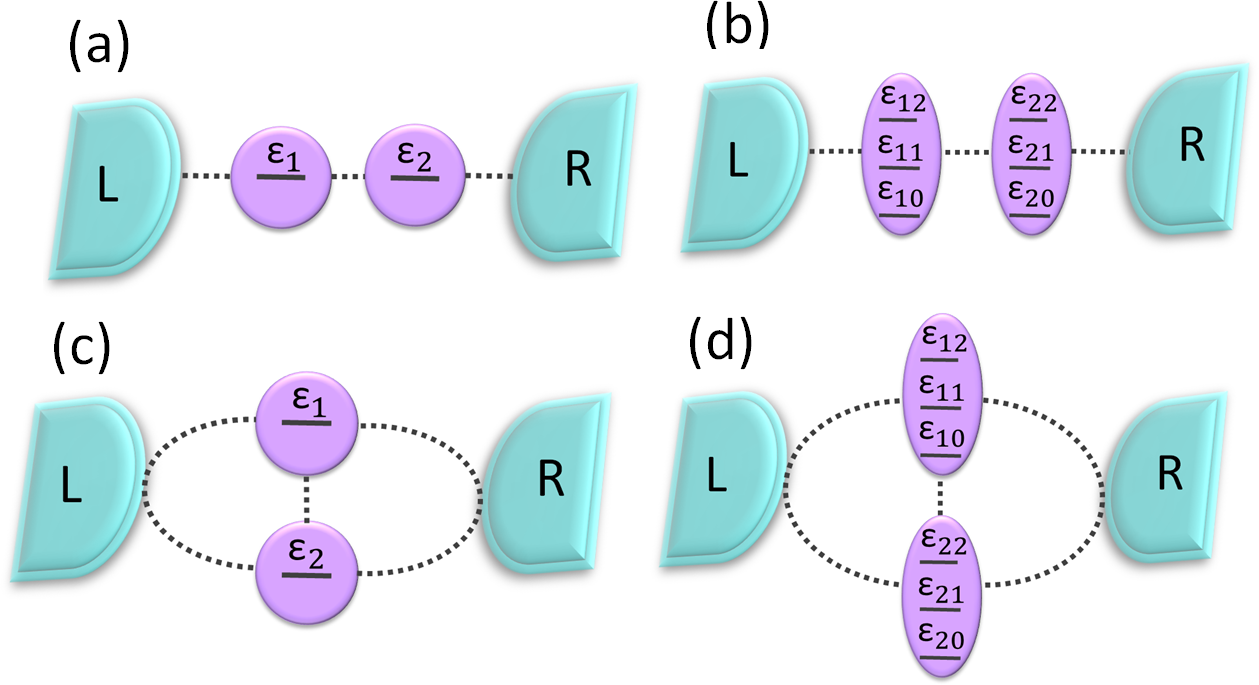}
\end{center}
\caption{The four studied geometries: (a)~SL-DQD in series, (b)~ML-DQD in series, (c)~SL-DQD in parallel, and (d)~ML-DQD in parallel. We take $\varepsilon_{1n}=\varepsilon_1+n\Delta\varepsilon_1$ and $\varepsilon_{2n}=\varepsilon_2+n\Delta\varepsilon_2$ with the integer index $n\in[0,2]$ in panels~(b) and (d) since one considers three energy levels in each of the two dots in the ML-DQD case. The dotted black lines symbolize the couplings between the various parts of the system.}
\label{fig2}
\end{figure}

{ In this section} one has analytically derived all the ingredients needed to characterize the electrical and thermoelectrical properties of a DQD whether it is in serial or parallel geometry. In the next two sections, one successively considers the DQD in series and in parallel for both single-level (SL) and multilevel (ML) dots. The electrical conductance $G=dI_L/dV$, with $V=(\mu_L-\mu_R)/e$ the bias voltage between the two leads, and the Seebeck coefficient $S=G^{-1}dI_L/d{\Delta}T$, with $\Delta T$ the temperature difference between the left and right leads\cite{Crepieux2015}, are calculated numerically with the help of Eq.~(\ref{I_final_maintext}). The zero-frequency charge susceptibility $\chi_c(0)$ and the total DQD occupancy $N= \langle\widehat N_1\rangle+ \langle\widehat N_2\rangle$ are calculated from  Eqs.~(\ref{exp_chi_derivative_new}) and (\ref{def_N}). The calculations are performed in the linear response regime, i.e. in the limit $V\rightarrow 0$ and $\Delta T\rightarrow 0$, but can be readily extended to the nonlinear response regime. We study the variations of $G$, $S$, $N$, and $\chi_c(0)$ as a function of gate voltages which act on the positions of energy levels  $\varepsilon_{1n}$ and  $\varepsilon_{2n}$ of the two dots constituting the DQD system.


\section{Discussion for a DQD in series}\label{section_DQD_series}

\subsection{SL-DQD in series}\label{section_SL_DQD_series}

\begin{figure}[t]
\hspace*{-0.3cm}\includegraphics[scale=.35]{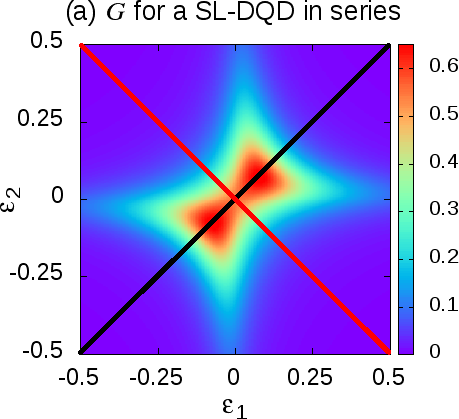}
\includegraphics[scale=.35]{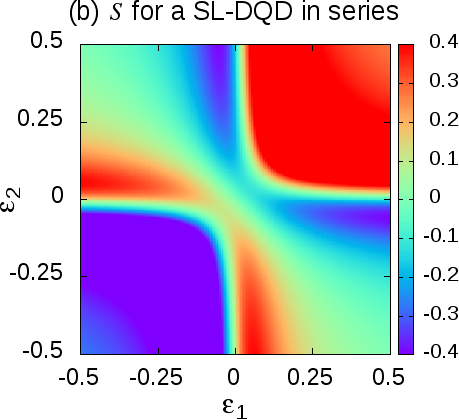}
\hspace*{-0.3cm}\includegraphics[scale=.35]{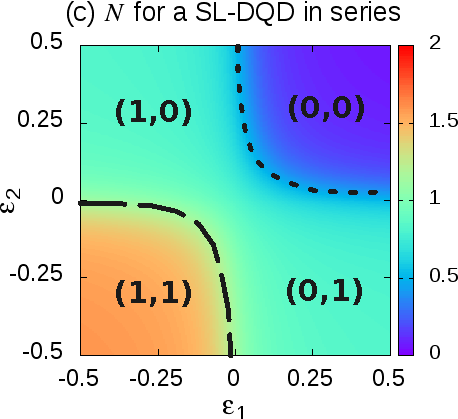}
\includegraphics[scale=.35]{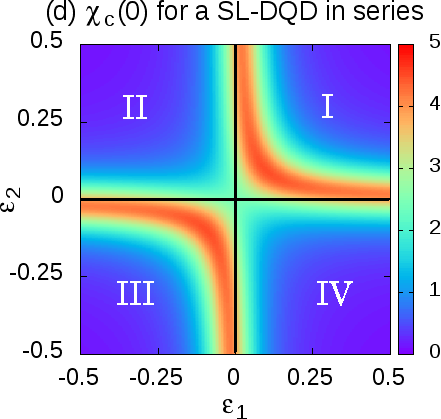}
\caption{Color-scale plots of (a)~the linear conductance~$G$, (b)~the Seebeck coefficient~$S$, (c)~the total dot occupancy~$N$, and (d)~the zero-frequency charge susceptibility~$\chi_c(0)$ for a SL-DQD connected in series as a function of $\varepsilon_1$ and $\varepsilon_2$ for $\mu_{L,R}=0$, $k_BT=0.01$, $\m{V}_{12}=0.1$, and $\Gamma_{L,11}=\Gamma_{R,22}=0.1$. In panel~(a) the black line marks the first diagonal of equation $\varepsilon_1=\varepsilon_2$, whereas the red line shows the second diagonal of equation $\varepsilon_1=-\varepsilon_2$ along which the plots of Fig.~\ref{figure_SL_cross-section_series} are drawn. In panel~(c) the domain with occupancies $ \langle\widehat N_1\rangle$ and $ \langle\widehat N_2\rangle$ of the dots 1 and 2 are indicated under the form $( \langle\widehat N_1\rangle, \langle\widehat N_2\rangle)$, and the dashed and dotted black arcs represent the boundary lines $\mathcal{B}_+$ and $\mathcal{B}_-$ between domains with different occupancies. Panel (d) shows the four quadrants: I~(top right), II~(top left), III~(bottom left) and IV~(bottom right).}
\label{figure_SL_series}
\end{figure}

\begin{figure}[t]
\hspace*{-0.3cm}\includegraphics[scale=.35]{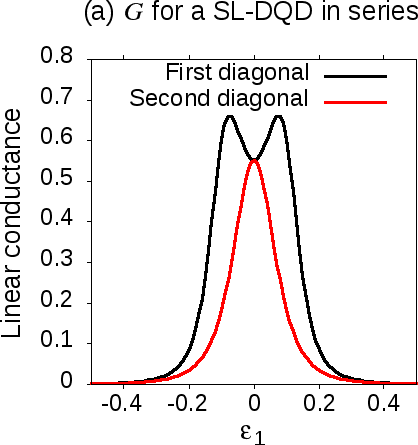}
\hspace*{0.2cm}\includegraphics[scale=.35]{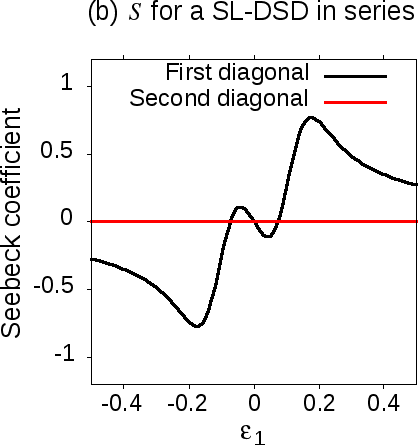}
\hspace*{-0.3cm}\includegraphics[scale=.35]{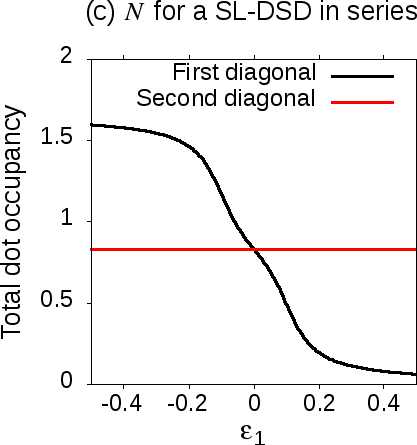}
\hspace*{0.2cm}\includegraphics[scale=.35]{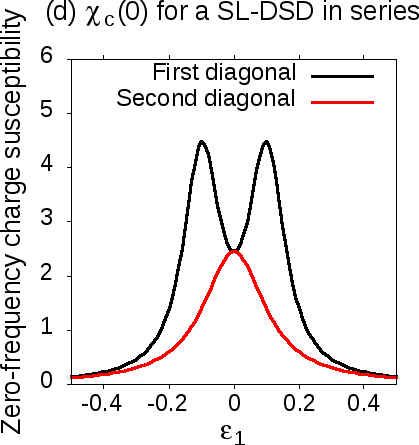}
\caption{Dependences as a function of $\varepsilon_1$ along the first diagonal $\varepsilon_1=\varepsilon_2$ and the second diagonal $\varepsilon_1=-\varepsilon_2$ of (a)~the linear electrical conductance~$G$, (b)~the Seebeck coefficient $S$, (c)~the dot occupancy $N$, and (d)~the zero-frequency charge susceptibility $\chi_c(0)$ for a SL-DQD connected in series. The choice of parameters is the same as in Fig.~\ref{figure_SL_series}}
\label{figure_SL_cross-section_series}
\end{figure}

We first consider the case of a DQD system in series with a single energy level in each of the two dots (see Fig.~\ref{fig2}(a)). Figure~\ref{figure_SL_series} shows the color-scale plots of $G$, $S$, $N$, and $\chi_c(0)$ as a function of the energies $\varepsilon_1$ and $\varepsilon_2$, whereas Fig.~\ref{figure_SL_cross-section_series} shows the plots of the same physical quantities as a function of $\varepsilon_1$ along either the first or the second diagonal of equation $\varepsilon_1=\varepsilon_2$ or $\varepsilon_1=-\varepsilon_2$, respectively. We describe the results obtained in Figs.~\ref{figure_SL_series} and \ref{figure_SL_cross-section_series} and then provide an interpretation for them. In a general way, we point out that all the color-scale plots in Fig.~\ref{figure_SL_series} have two axes of symmetries which are the first and second diagonals. Figure~\ref{figure_SL_series}(a) shows that the conductance $G$ is the largest in the central region surrounding the origin point O ($\varepsilon_1=0, \varepsilon_2=0$), with the presence of two peaks along the first diagonal, equidistant from O. Besides, O behaves as a saddle point in the sense that $G$ is maximal at that point when sweeping along the second diagonal direction, while it is a local minimum along the first diagonal one (see Fig.~\ref{figure_SL_cross-section_series}(a)). When getting farther from the origin O, $G$ gradually decreases forming a star-shaped pattern in the plane $(\varepsilon_1, \varepsilon_2)$ as displayed in blue color in Fig.~\ref{figure_SL_series}(a), until reaching the zero value in the remaining parts of the plane. The color-scale plot of the Seebeck coefficient displayed in Fig.~\ref{figure_SL_series}(b) shows that $S$ is zero (green color) along the boundary lines $\mathcal{B}_+$ and $\mathcal{B}_-$ located in quadrants III and I, as well as inside a band located on either side of the second diagonal. It takes positive values (red color) inside the top-right domain delimited by the boundary line $\m{B}_-$, and vice-versa negative values (violet color) inside the respective bottom-left domain delimited by $\m{B}_+$. In the intermediate area located between the two boundary lines, $S$ exhibits pockets of local maxima and minima located in the vicinity of $\mathcal{B}_+$ and $\mathcal{B}_-$. In the plot of $S$ as a function of $\varepsilon_1$ along the first and second diagonal displayed in Fig.~\ref{figure_SL_cross-section_series}(b), one can see that $S$ is zero all along the second diagonal, while it changes of sign three times when sweeping along the first diagonal, once at $\varepsilon_1=0$ and the two other times at the positions of the maxima of $G$ observed in Fig.~\ref{figure_SL_cross-section_series}(a). The behavior for $S$ which we obtain with these three observed changes of sign, are in agreement with {  both the theoretical results of Ref.~\onlinecite{Chen2000} and} the experimental results obtained in GaAs/AlGaAs heterostructures which are reported in Fig.~3 of Ref.~\onlinecite{Thierschmann2013} and Fig.~4 in Ref.~\onlinecite{Thierschmann2016}. We underline that in the experimental works $V_\text{th} = -S\Delta T$ is plotted instead of $S$ in the results presented here. Figure~\ref{figure_SL_series}(c) shows the evolution of $N$ as a function of $\varepsilon_1$ and $\varepsilon_2$. It reveals the charge stability diagram with the presence of four domains denoted as $( \langle\widehat N_1\rangle, \langle\widehat N_2\rangle)$ inside which $ \langle\widehat N_1\rangle$ and $ \langle\widehat N_2\rangle$ take values close to (0,0), (1,0), (0,1) or (1,1). As can be seen, the (0,0) and (1,1) domains are shrunk within the I and III quadrants, compared to the uncoupled DQD case (at $\mathcal{V}_{12}=0$) for which the diagram would have shown a tiling on a square lattice set on the four quadrants (see Figs.~\ref{figure_VP}(a), (c), and (e) in Appendix~\ref{appendix_Hdot}). Note that the boundary lines $\m{B}_-$ and $\m{B}_+$ delimiting the (0,0) and (1,1) domains have the shape of arcs and that $N$ takes a constant value along the second diagonal (see red curve in Fig.~\ref{figure_SL_cross-section_series}(c)). The 
evolution of the zero-frequency charge susceptibility $\chi_c(0)$ displayed in Figs.~\ref{figure_SL_series}(d) and \ref{figure_SL_cross-section_series}(d) follows the same trend. It shows the presence of two lines of maxima, located precisely on the boundary lines  $\mathcal{B}_+$ and $\mathcal{B}_-$ highlighted in Fig.~\ref{figure_SL_series}(c). As for $G$, the origin O behaves as a saddle point, $\chi_c(0)$ being maximal when sweeping along the second diagonal direction, while it is a local minimum along the first diagonal one. 

The results obtained for $G$, $S$, $N$ and $\chi_c(0)$ can be interpreted in a simple way by relying on the properties presented in Appendix~\ref{appendix_Hdot}. It is explained how, when the DQD system is disconnected from the leads, the hamiltonian $\m{\widehat H}_\text{dots}$ describing the system can be diagonalized leading to the eigenenergies $E^+_\text{dots}$ and $E^-_\text{dots}$, the values of which are given by Eq.~(\ref{eig_2}).{  The corresponding antibonding and bonding eigenstates\cite{VanDerWiel,Guevara2003} $|+\rangle$ and $|-\rangle$ are defined by Eqs.~(\ref{def_EV1}) and (\ref{def_EV2})}. Consequently the spectral density $A_{11}(\varepsilon)$ in the dot 1, respectively $A_{22}(\varepsilon)$ in the dot 2, is a linear combination of delta functions within a multiplicative factor $2\pi$, centered at the values of eigenenergies  $E^+_\text{dots}$ and  $E^-_\text{dots}$, corresponding to the spectral densities of eigenstates $A_{+}(\varepsilon)$ and $A_{-}(\varepsilon) $, with weighting factors equal to $|u|^2$ and $|v|^2$, respectively $|v|^2$  and $|u|^2$  in the dot 2, where $|u|^2$ and $|v|^2$ are defined by Eqs.~(\ref{def_u}) and (\ref{def_v}). The physical meaning of this diagonalization is that the DQD system behaves as an effective single quantum dot with two energy levels at energies $E^-_\text{dots}$ and  $E^+_\text{dots}$. The charge stability diagram of the disconnected DQD system can easily be derived from the latter results. At equilibrium the boundary lines between the domains of different occupations are obtained when any of the two levels of energies $E^+_\text{dots}$ and  $E^-_\text{dots}$ is aligned with the chemical potential of the leads, taken to 0 ($\mu_{L,R}=0$). The equations of the boundary lines $\m{B}_+$ and $\m{B}_-$, hence given by  $E^+_\text{dots}=0$ and  $E^-_\text{dots}=0$, are $\varepsilon_1\varepsilon_2=|\m{V}_{12}|^2$ (see Appendix \ref{appendix_Hdot}). They correspond to the two branches of an hyperbole as shown in Figs.~\ref{figure_VP}(a), (c), and (e) for different values of the interdot coupling $\m{V}_{12}$. The distance between the two branches is minimal along the first diagonal, taking the value of $2|\m{V}_{12}|$. They correspond to the two boundary lines found in quadrants I and III in Fig.~\ref{figure_SL_series}(c): the curve $\m{B}_-$ of equation $E^-_\text{dots}= 0$ corresponds to the boundary line in quadrant I, whereas the curve $\m{B}_+$ of equation  $E^+_\text{dots}= 0$ corresponds to the boundary line in quadrant III. Inside the top-right domain delimited by the curve $\m{B}_-$, both $E^-_\text{dots}$ and $E^+_\text{dots}$ are positive and the two corresponding energy levels are empty. It gives rise to the domain (0,0) in the charge stability diagram (by making use of the results on the spectral densities $A_{ii}(\varepsilon)$ mentioned above and by noticing that $|u|^2+|v|^2=1$). Inside the bottom-left domain delimited by the curve $\m{B}_+$, both $E^-_\text{dots}$ and $E^+_\text{dots}$ are negative and the corresponding two energy levels are occupied. It gives rise to the domain (1,1) in the charge stability diagram. Inside the area located between the two boundary lines $\m{B}_+$ and $\m{B}_-$, centered around the second diagonal, $E^-_\text{dots}$ is negative while $E^+_\text{dots}$ is positive, hence only the lower energy level at $E^-_\text{dots}$ is occupied. It corresponds either to the domain (1,0) or (0,1) in the charge stability diagram. The line of separation between these latter two domains is along the first diagonal. The results obtained above from the diagonalization of  $\m{\widehat H}_\text{dots}$ apply to the case of a DQD disconnected from the leads. However it is easy to realize that connecting the DQD system to the leads, would introduce a broadening of the eigenenergy levels described above together with an eventual renormalization of the eigenenergies values. It would not change the general shape of the charge stability diagram discussed above, but would simply widen the boundary lines separating the different domains. It allows one to explain the charge stability 
diagram obtained in Fig.~\ref{figure_SL_series}(c) with a remarkable agreement on the value 
of the minimal distance observed between the two boundary lines, equal to~$2|\m{V}_{12}|$.

We now interpret the results obtained for the linear electrical conductance as reported in Figs.~\ref{figure_SL_series}(a) and \ref{figure_SL_cross-section_series}(a). From Eqs.~(\ref{I_final}) and (\ref{def_EV1}-\ref{def_v}), the linear conductance at zero temperature is proportional to: $\Gamma_{L,11}\Gamma_{R,22} |uv|^2 [A_+(0)+ A_-(0)] $. The spectral densities $A_+(0)$ and $A_-(0)$ are maximal when the point $(\varepsilon_1, \varepsilon_2)$ corresponding to the energies of the dots falls in one of the boundary lines $\m{B}_+$ and $\m{B}_-$. However in order to get the linear conductance, one has to weight the result for the spectral densities of eigenstate at $\varepsilon=0$ by a multiplicative coherence factor equal to  $|uv|^2$. Typically $|uv|^2$ is the largest along the first diagonal $(\varepsilon_1=\varepsilon_2)$ where it equals the value 1/4. Moreover $u\rightarrow 1$ and $v\rightarrow 0$ along the end-part of the boundary lines asymptotic to the $\varepsilon_2$-axis, whereas $u\rightarrow 0$ and $v\rightarrow 1$ along the end-part of the boundary lines asymptotic to the $\varepsilon_1$-axis. Combining these arguments, the peaks of $G$ in the plane $(\varepsilon_1, \varepsilon_2)$ arise at the intersection of the boundary lines $\m{B}_+$ and $\m{B}_-$ and of the first diagonal as seen in Figs.~\ref{figure_SL_series}(a) and \ref{figure_SL_cross-section_series}(a). 
The results obtained for the Seebeck coefficient in Figs.~\ref{figure_SL_series}(a) and \ref{figure_SL_cross-section_series}(a) can be interpreted in the same way. From Ref.~\onlinecite{Turek2002}, the Seebeck coefficient is proportional to the average energy of the charge carriers: $S=-\langle E-\mu\rangle /k_BT$ which is zero in two types of situations: either when the chemical potential $\mu$ (with $\mu$ taken to 0 here) falls at the center of one of the peaks of the spectral density of states, that is to say at $E^+_\text{dots}$ and $E^-_\text{dots}$, or when the chemical potential falls at equal distance from the two peaks. In either case, there are as many carriers with negative energies than carriers with positive energies and hence the average energy of carriers is zero (electron-hole symmetry). This explains why the zeros of $S$ in Fig.~\ref{figure_SL_series}(b) occur along the boundary lines $\m{B}_-$ and $\m{B}_+$, as well as along the second diagonal, where $(E^+_\text{dots} + E^-_\text{dots})/2=0$ since $\varepsilon_1=-\varepsilon_2$ there, as seen in Figs.~\ref{figure_SL_cross-section_series}(b). {  The cancellation of $S$ could be used to measure the value of the interdot coupling $\mathcal{V}_{12}$ which is directly given by the distance between the canceling points.} The interpretation of the results obtained for the zero-frequency charge susceptibility follows in the same way: $\chi(0)$ determined from Eq.~(\ref{exp_chi_derivative_new}) is maximal in the plane $(\varepsilon_1, \varepsilon_2)$ when the spectral densities $A_+(0)$ and $A_-(0)$ are maximal, i.e. when the point $(\varepsilon_1, \varepsilon_2)$ falls in one of the boundary lines $\m{B}_+$ and~$\m{B}_-$.


\subsection{ML-DQD in series}\label{section_ML_DQD_series}

\begin{figure}[t]
\hspace*{-0.3cm}\includegraphics[scale=.35]{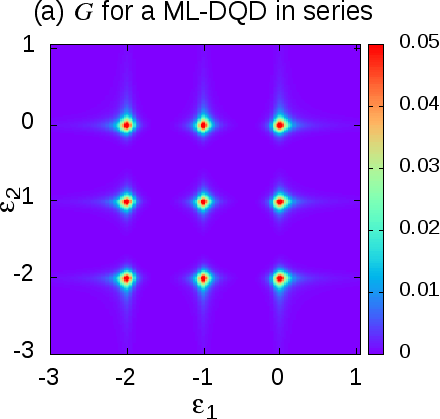}
\includegraphics[scale=.35]{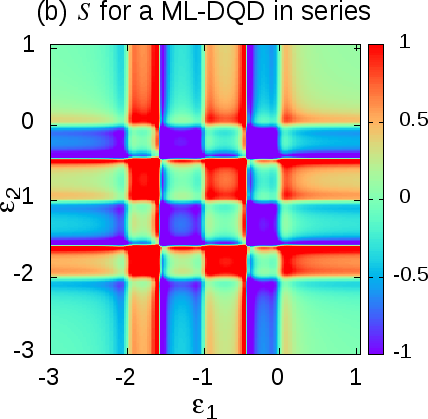}
\hspace*{-0.3cm}\includegraphics[scale=.35]{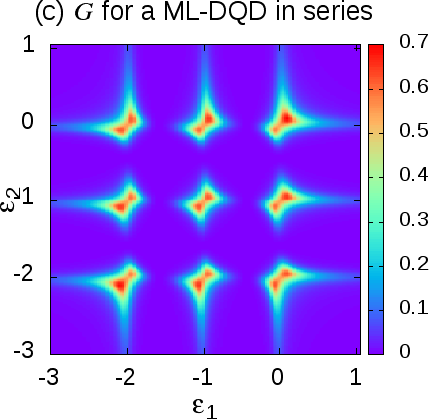}
\includegraphics[scale=.35]{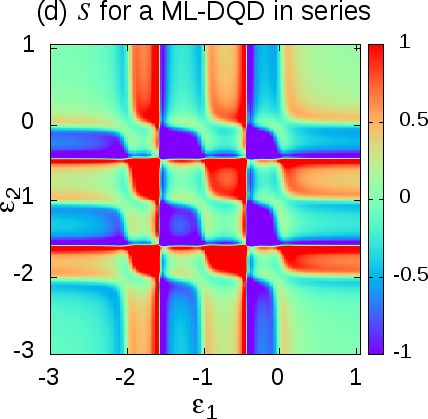}
\hspace*{-0.3cm}\includegraphics[scale=.35]{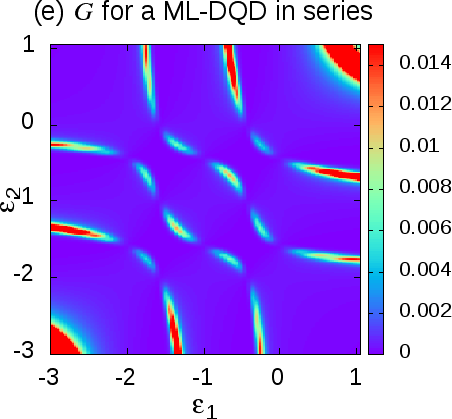}
\includegraphics[scale=.35]{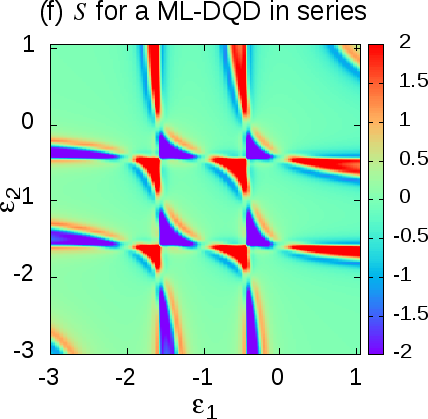}
\caption{Color-scale plots of $G$ and $S$ for a ML-DQD connected in series as a function of $\varepsilon_1$ and $\varepsilon_2$ at $\mu_{L,R}=0$, $k_BT=0.01$  for: (a), (b)~weak interdot coupling ($\m{V}_{12}=0.01$, $\Gamma_{L,11}=\Gamma_{R,22}=0.1$); (c), (d)~intermediate interdot coupling  ($\m{V}_{12}=\Gamma_{L,11}=\Gamma_{R,22}=0.1$); and (e), (f)~strong interdot coupling ($\m{V}_{12}=0.5$, $\Gamma_{L,11}=\Gamma_{R,22}=0.01$).}
\label{figure_G_1}
\end{figure}

\begin{figure}[t]
\hspace*{-0.3cm}\includegraphics[scale=.35]{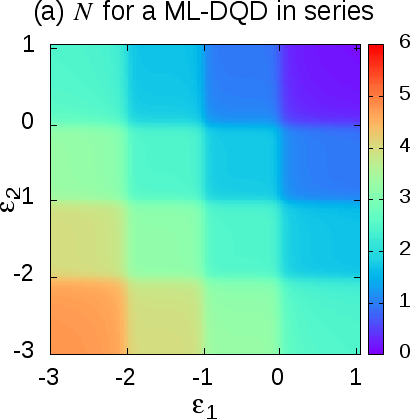}
\includegraphics[scale=.35]{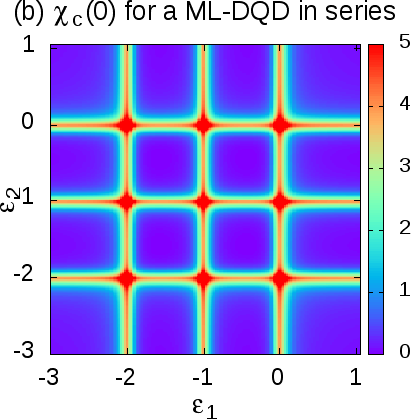}
\hspace*{-0.3cm}\includegraphics[scale=.35]{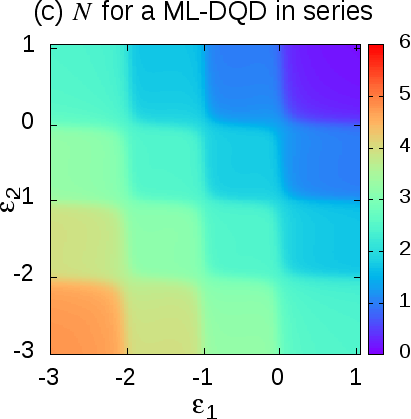}
\includegraphics[scale=.35]{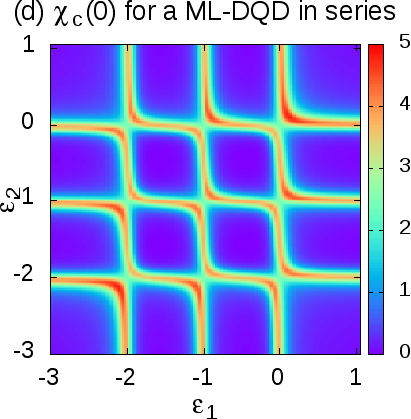}
\hspace*{-0.3cm}\includegraphics[scale=.35]{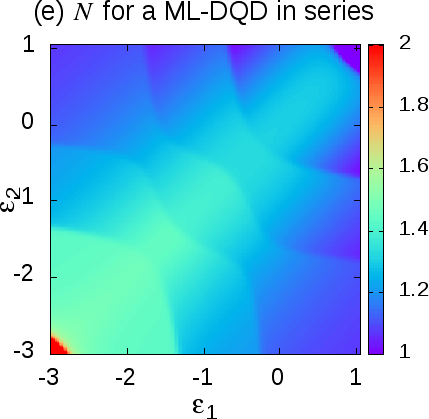}
\includegraphics[scale=.35]{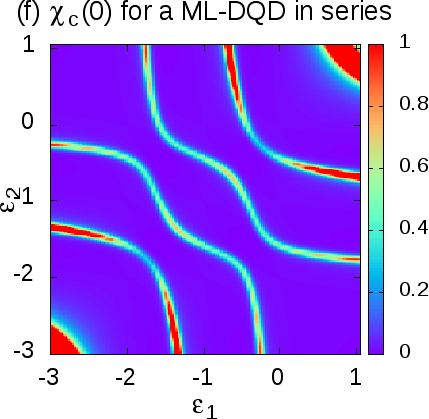}
\caption{Color-scale plots of $N$ and $\chi_c(0)$ for a ML-DQD connected in series as a function of $\varepsilon_1$ and $\varepsilon_2$.  The parameters are the same as in Fig.~\ref{figure_G_1}.}
\label{figure_S_1}
\end{figure}

We examine the case of a ML-DQQ in series. As an example we consider the situation where each of the quantum dots $i=1,2$ has three energy levels of energies $\varepsilon_{in}=\varepsilon_i+n\Delta \varepsilon_i$ with the integer index $n\in[0,2]$ and $\Delta \varepsilon_i=1$ (see Fig.~\ref{fig2}(b)). The results obtained for the color-scale plots of $G$, $S$ and $N$, $\chi_c(0)$ as a function of  $\varepsilon_1$ and $\varepsilon_2$ are reported in Figs.~\ref{figure_G_1} and \ref{figure_S_1} respectively in the different interdot coupling regimes.

In the weak interdot coupling regime, i.e. for $|\m{V}_{12}| \ll \Gamma_{L,11}, \Gamma_{R,22}$, the electrical conductance $G$ displayed in Fig.~\ref{figure_G_1}(a) exhibits peaks centered at the nodes of a square lattice constituted by the vertical lines $\varepsilon_{1n}=0$ and horizontal lines $\varepsilon_{2n}=0$. Moreover in the continuation of these peaks, one glimpses a slight enhancement of $G$ along the lines of the square lattice. The results for $N$ reported in Fig.~\ref{figure_S_1}(a) reveals the charge stability diagram with the presence of $4\times 4=16$ domains where $N$ changes by plateau. The boundary lines separating the domains coincide with the lines of the square lattice highlighted above. The top-right corner domain corresponds to the completely empty DQD system denoted as (0,0). Whereas one would have expected a completely filled DQD system in the bottom-left corner domain with an occupation (3,3), we point out that the maximal value of $N$ reached there is close to 5, instead of 6, due to the relatively weak value of the dot-lead couplings ($\Gamma_{L,11}=\Gamma_{R,22}= 0.1$). In the color-scale plot of  $\chi_c(0)$ shown in Fig.~\ref{figure_S_1}(b),  $\chi_c(0)$ is maximal along the same boundary lines as for $N$, reaching the zero value inside the delimited domains. Finally the color-scale plot of $S$ displayed in Fig.~\ref{figure_G_1}(b) shows that $S$ changes of sign several times in the plane $(\varepsilon_1,\varepsilon_2)$ {  as previously highlighted in Ref.~\onlinecite{Wierzbicki2010}}.

In the intermediate interdot regime, i.e. for $|\m{V}_{12}|$ of the order of $\Gamma_{L,11}, \Gamma_{R,22}$, the results for the color-scale plots of $G$ displayed in Fig.~\ref{figure_G_1}(c) are strongly reminiscent of the results obtained in the case of the single-level DQD in series (see Fig.~\ref{figure_SL_series}(a)). $G$ is the largest in the central areas surrounding the nodes of the square lattice, with the presence of two peaks on either side of the nodes. These two peaks are located along a line parallel to the first diagonal. When getting farther from these nodes, $G$ gradually decreases along lines forming a characteristic star-shaped pattern, whereas it is zero in the remaining part of the plane. The results for $N$ displayed in Fig.~\ref{figure_S_1}(c) still show the presence of 16 domains of different occupancies. The boundary lines between these domains are no longer straight lines but becomes sinuous, the vertices of the square lattice having separated into two triple points {  in complete agreement with standard results\cite{VanDerWiel,Hanson2007}}. These triple points are at the intersection of the boundaries delimiting three domains of different occupancies. In the color-scale plot of $\chi_c(0)$ shown in Fig.~\ref{figure_S_1}(d), $\chi_c(0)$ is maximal along the same boundary lines as in Fig.~\ref{figure_S_1}(c), reaching the zero value inside the delimited domains. Finally the color-scale plot of $S$ displayed in Fig.~\ref{figure_G_1}(d) shows that $S$ exhibits successive changes of sign in the plane $(\varepsilon_1,\varepsilon_2)$, being of positive sign in the areas delimited by the convex parts of the sinuous boundary line facing top-right, i.e. similar to the area in quadrant I of Fig.~\ref{figure_SL_series}(b), and of negative sign in its concave parts facing bottom-left, i.e., similar to the area in quadrant III in Fig.~\ref{figure_SL_series}(b). Besides $S$ is zero inside broad strips surrounding either the second diagonal or the two lines parallel to the second diagonal. Hence $S$ changes sign nine times when sweeping along the first diagonal, instead of five times in Fig.~\ref{figure_G_1}(b). 

In the strong interdot coupling regime, i.e. for $|\m{V}_{12}| \gg \Gamma_{L,11}, \Gamma_{R,22}$, the square lattice structure visible in the previous figures has disappeared, giving place to an oblique structure in the direction of the second diagonal as can be seen in Figs.~\ref{figure_G_1}(e) and \ref{figure_G_1}(f), and Figs.~\ref{figure_S_1}(e) and \ref{figure_S_1}(f) for $G$, $S$, $N$ and $\chi_c(0)$. This means that the two quantum dots have merged into one single quantum dot of occupation $\langle\widehat N_1\rangle+\langle\widehat N_2\rangle$. In Fig.~\ref{figure_S_1}(e), one sees that the change of $N$ by plateau observed in the weak and intermediate regimes, is replaced by a smooth variation. The orders of magnitude obtained for $N$ and $\chi_c(0)$ are ten times smaller in the case of strong coupling regime compared to the weak and intermediate coupling regimes, the observed reduction being of the same magnitude as the reduction of $\Gamma_{L,11}$ and $\Gamma_{R,22}$ values. The results for $S$ shown in Fig.~\ref{figure_G_1}(f) are in qualitative agreement with the experimental ones obtained in GaAs/AlGaAs heterostructures displayed in Fig.~2 of Ref.~\onlinecite{Thierschmann2013}. 

The whole set of these results can be physically interpreted by relying on the properties presented in Appendix~\ref{appendix_Hdot} where it is shown how the hamiltonian  $\m{\widehat H}_\text{dots}$ describing the three-level DQD system can be diagonalized leading to six eigenenergies $ E_\text{dots}^\lambda$ whose values are determined numerically. The charge stability diagram showing $N$ can easily be derived from that. As in the case of the SL-DQD system, the boundary lines between the domains of different occupations are obtained when any of six levels of energies $E_\text{dots}^\lambda$ is aligned with the chemical potentials of the leads both taken to 0 at equilibrium ($\mu_{L,R}=0$). The boundary lines of equations $E_\text{dots}^\lambda=0$ are precisely the  equienergetic curves represented in Figs.~\ref{figure_VP}(b), \ref{figure_VP}(d), and \ref{figure_VP}(f) for different values of the interdot coupling $\m{V}_{12}$. One recovers the shapes of the boundary lines obtained above, going from straight lines on a square lattice in the weak interdot coupling regime to sinuous lines in the intermediate and strong interdot coupling regime. As emphasized in Appendix~\ref{appendix_Hdot}, as soon as $\m{V}_{12}$ becomes finite, a level anticrossing effect takes place at the vicinity of the nodes of the square lattice, the vertices of the square lattice splitting into two triple points as observed in Fig.~\ref{figure_VP}. The interpretation of the other results obtained for $G$, $S$, and $\chi_c(0)$ is modeled on the one given above in the single-level case with the presence of some coherence factors which plays a role of extinction for some parts of the equienergetic curves $E_\text{dots}^\lambda=0$.

\section{Discussion for a DQD in parallel}\label{section_DQD_parallel}

\subsection{SL-DQD in parallel}\label{section_SL_DQD_parallel}

\begin{figure}[t]
\hspace*{-0.3cm}\includegraphics[scale=.35]{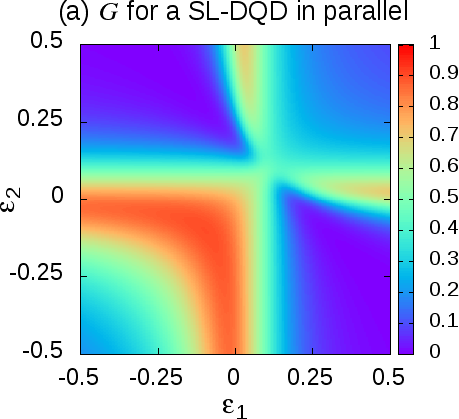}
\includegraphics[scale=.35]{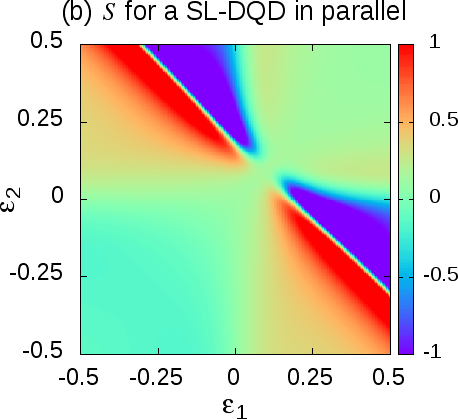}
\hspace*{-0.3cm}\includegraphics[scale=.35]{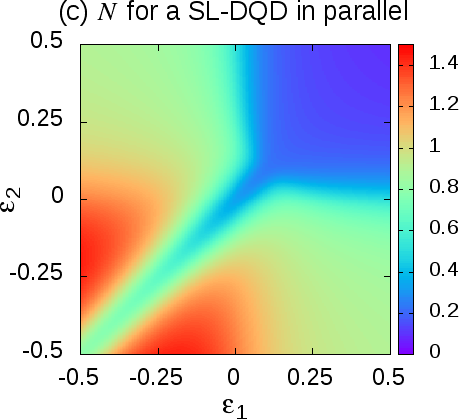}
\includegraphics[scale=.34]{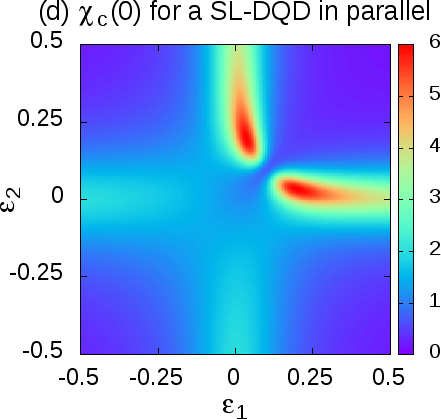}
\caption{Color-scale plots of (a)~the linear electrical conductance $G$, (b)~the Seebeck coefficient $S$, (c)~the total dot occupancy $N$, and (d)~the zero-frequency charge susceptibility $\chi_c(0)$ for a SL-DQD connected in parallel as a function of $\varepsilon_1$ and $\varepsilon_2$ for $\mu_{L,R}=0$, $k_BT=0.01$, $\m{V}_{12}=0.1$, and $\Gamma_{\alpha,ij}=0.1$ for both $\alpha=L,R$ and $i,j=1,2$. 
}
\label{figure_SL_parallel}
\end{figure}

\begin{figure}[t]
\hspace*{-0.3cm}\includegraphics[scale=.35]{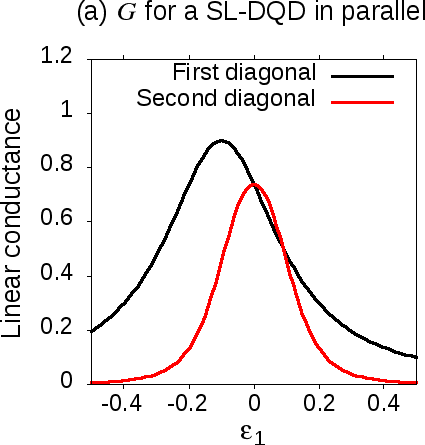}
\hspace*{0.2cm}\includegraphics[scale=.35]{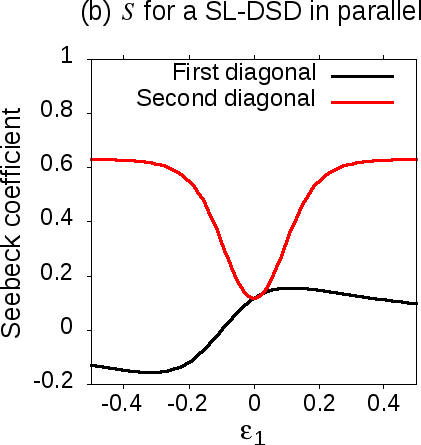}
\hspace*{-0.3cm}\includegraphics[scale=.35]{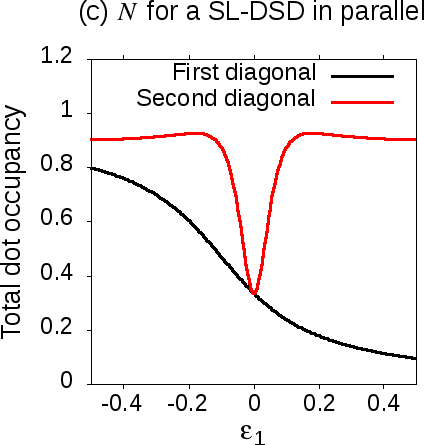}
\hspace*{0.2cm}\includegraphics[scale=.34]{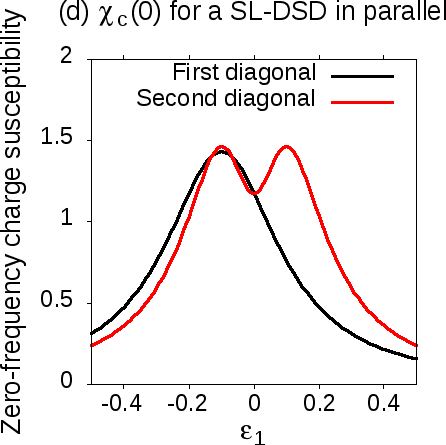}
\caption{Dependences as a function of $\varepsilon_1$ along the first diagonal $\varepsilon_1=\varepsilon_2$ and second diagonal $\varepsilon_1=-\varepsilon_2$ of (a)~$G$, (b)~$S$, (c)~$N$, and (d)~$\chi_c(0)$ for a SL-DQD connected in parallel. The parameters are the same as in Fig.~\ref{figure_SL_parallel}.}
\label{figure_SL_cross-section_parallel}
\end{figure}

The schematic diagram of the DQD in parallel with a single level in each dot is depicted in Fig.~\ref{fig2}(c). The color-scale plots for $G$, $S$, $N$ and $\chi_c(0)$ as a function of $\varepsilon_1$ and $\varepsilon_2$ are shown in Fig.~\ref{figure_SL_parallel}. The results are very different from those obtained in the case in series even if in all the plots in Fig.~\ref{figure_SL_parallel}, one still glimpses the presence of the boundary lines $\m{B}_+$ and $\m{B}_-$ in places almost unchanged compared to the case in series. Strikingly, whereas the color-scale plots of Fig.~\ref{figure_SL_series} obtained in the serial cases had two axes of symmetry along the first and the second diagonals, only the axial symmetry with respect to the first diagonal is conserved in Fig.~\ref{figure_SL_parallel} while the one with respect to the second diagonal is lost. The color-scale of $G$ in Fig.~\ref{figure_SL_parallel}(a) shows that instead of the two peaks for $G$ in the serial case which occurred at the intersection of the boundary lines  $\m{B}_+$ and $\m{B}_-$ and of the first diagonal, the positions of the maxima of $G$ now spread all along the boundary line $\m{B}_+$ in the bottom-left corner, whereas those in the top-right corner are located along the end-parts of the boundary line $\m{B}_-$. Moreover one can notice that the conductance ridges thus formed are much broader in the bottom-left corner than in the top-right corner, with higher values reached along the former rather than along the latter ones. We also point out that the amplitude of the conductance to the maximum is about 40$\%$ higher than in the serial case. The plots of $G$ along the first and second diagonals displayed in Fig.~\ref{figure_SL_cross-section_parallel}(a) bring a complementary information to that. When sweeping along the first diagonal, $G$ exhibits a single peak at a negative value of $\varepsilon_1$ { in perfect agreement with the result displayed in Fig.~2 of Ref.~\onlinecite{Guevara2003}}, whereas the peak of $G$ along the second diagonal is centered at the zero value. 

The results obtained for $S$ displayed in Fig.~\ref{figure_SL_parallel}(b) shows that $S$ still takes positive values inside the top-right domain and negative values inside the bottom-left domain as it was the case for the single-level DQD case, with a strong reduction in the order of magnitude of the amplitude compared to the serial configuration results shown in Fig.~\ref{figure_SL_series}(b). In the intermediate area located between these two domains, $S$ exhibits a series of minima of negative sign inside a pair of two triangular pockets along the boundary line delimiting the top-right domain as in Fig.~\ref{figure_SL_series}(b). Nevertheless we point out three major differences: (i) the base of these triangles is a straight line parallel to the second diagonal shifted to the top-right corner from the second diagonal by a distance equal to $|\m{V}_{12}|$ and one has $S=0$ on this base; (ii) the two triangles are disjoint at the center showing a gap around the first diagonal in which $S$ takes a positive value close to zero; and (iii) the negative values reached inside the triangles are about twice larger than inside the bottom-left corner. Moreover one still observes the presence a series of maxima of positive sign inside a pair of two triangular pockets as in Fig.~\ref{figure_SL_series}(b), the difference in the parallel case is that these two triangular pockets are contiguous to the previous ones.  Besides they share with the negative sign triangles the same three peculiarities i.e. (i), (ii), and (iii) as far as we speak of their positive sign values. Finally $S$ gradually decreases keeping a positive sign inside the area located between the latter positive sign triangles and the boundary line surrounding the bottom-left domain, exhibiting a gap around the first diagonal. The plots of $S$ along the first and second diagonals displayed in Fig.~\ref{figure_SL_cross-section_parallel}(b) completes this information. Along the first diagonal, $S$ goes from negative to positive values with increasing $\varepsilon_1$, showing one change of sign instead of three in the serial configuration. Along the second diagonal, $S$ is no longer zero since the electron-hole symmetry holding in the serial case is now lost, instead $S$ keeps a positive value all along the second diagonal with a marked minimum around $\varepsilon_1=0$. The electron-hole symmetry is however restored with $S=0$ on a line which corresponds to the base of the triangles discussed there before (see the green line parallel to the second diagonal in Fig.~\ref{figure_SL_parallel}(b)).

The color-scale plot of $N$ displayed in Fig.~\ref{figure_SL_parallel}(c) still shows the presence of the boundary lines in top-right and bottom-left corners as in Fig.~\ref{figure_SL_series}(c). However one can notice important differences in comparison to the results obtained for the case in series : (i) the value of $N$ is strongly reduced inside the domain in the bottom-left corner, reaching the value 0.8 instead of 1.6; (ii) an elongated tip is formed along the first diagonal extending from the domain in the top-right corner to the other side where it cuts the bottom-left corner domain in half. Strikingly the value of $N$ is strongly reduced along this tip going from 0 to 0.8 when sweeping along the first diagonal as shown in Fig.~\ref{figure_SL_cross-section_parallel}(c). Moreover the plot of $N$ along the second diagonal displayed in Fig.~\ref{figure_SL_cross-section_parallel}(c) too shows a marked minimum at $\varepsilon_1=0$ at the crossing with the elongated tip previously reported.

Finally the evolution of $\chi_c(0)$ displayed in Fig.~\ref{figure_SL_parallel}(d) follows the same trend. It shows lines of maxima along the same boundary lines highlighted in Fig.~\ref{figure_SL_parallel}(a) but the maxima are much more pronounced along the boundary lines in the top-right corner in comparison to the ones in the bottom-left corner. Besides these two lines of maxima are cut into half at the crossing with the first diagonal with the opening of a gap around it. As shown in Fig.~\ref{figure_SL_cross-section_parallel}(d), when sweeping along the first diagonal, $\chi_c(0)$ exhibits a peak at a negative sign value of $\varepsilon_1$, whereas $\chi_c(0)$ along the second diagonal exhibits two peaks located on both sides of the zero value.

\begin{figure}
\begin{center}
\hspace*{-0.3cm}\includegraphics[scale=.35]{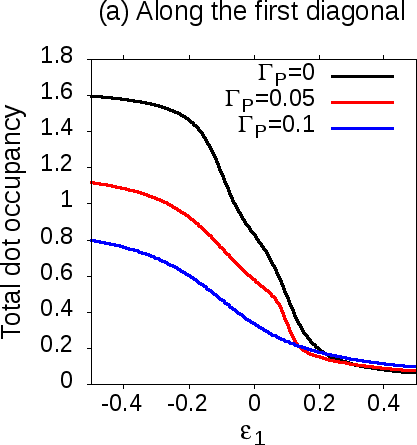}
\includegraphics[scale=.35]{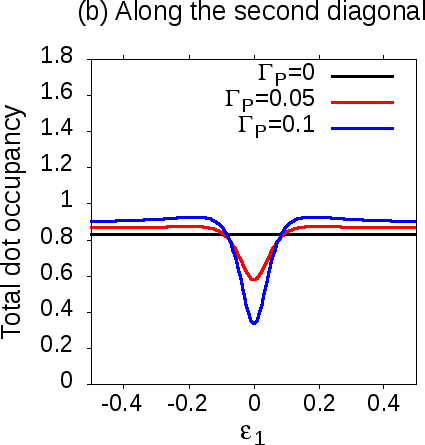}
\end{center}
\caption{Dependences as a function of $\varepsilon_1$ of $N$ along (a)~the first diagonal and (b)~the second diagonal as a function of $\varepsilon_1$ at $\mu_{L,R}=0$, $k_BT=0.01$, $\m{V}_{12}=0.1$ for $\Gamma_{L,11}=\Gamma_{R,22}=0.1$, all the other dot-lead couplings being equal to $\Gamma_P$, meaning that the SL-DQD is in series when $\Gamma_P=0$ (black curves) and in parallel when $\Gamma_P=0.1$ (blue curves). The red curves correspond to an intermediate situation ($\Gamma_P=0.05$).}
\label{figure_N}
\end{figure}

As far as the physical interpretation of the results obtained for $G$, $S$, $N$, and $\chi_c(0)$ is concerned, we would say that even though the basic feature comes from the formation of the boundary lines which occurs when one of the eigenenergies in the system is aligned with the chemical potential of the leads with the coherence factors acting as extinction factors, as explained in Sec.~\ref{section_SL_DQD_series}, the parallel configuration introduces some noticeable changes with comparison to the serial configuration as the existence of more than one transmission channels. The differences observed in the behaviors of $G$, $S$, $N$, and $\chi_c(0)$ compared to the case in series results from the interference effects which take place in the presence of two transmission channels as it is the case in the parallel configuration. We give below some simple arguments, developed within the zero temperature and zero interdot coupling limit, which help in elucidating the origin of the formation of the elongated tip mentioned above. In the limit of zero temperature and zero interdot coupling, the expression of $N$ can be derived analytically from Eq.~(\ref{def_N}). One gets for a SL-DQD in series
\begin{eqnarray}\label{N_series}
N_\text{series}=1-\frac{1}{\pi}\arctan\left(\frac{2\varepsilon_1}{\Gamma_{L,11}}\right)-\frac{1}{\pi}\arctan\left(\frac{2\varepsilon_2}{\Gamma_{R,22}}\right)\nonumber\\
\end{eqnarray}
This result explains why according to the sign of $\varepsilon_1$ and $\varepsilon_2$, the total occupancy $N$ varies by plateau along which it takes either the value 0, 1 or 2 at most, with a change from one plateau to the other spreading over a width $\Gamma_{L,11}$ or $\Gamma_{R,22}$. For a SL-DQD in parallel, one gets when $\varepsilon_1=\varepsilon_2$,
\begin{eqnarray}\label{N_parallel}
N_\text{parallel}=\frac{1}{2}-\frac{1}{\pi}\arctan\left(\frac{\varepsilon_1}{2\Gamma}\right)
\end{eqnarray}
where $\Gamma=\Gamma_{\alpha,ij}$ for any $\alpha=L,R$ and $i,j=1,2$, i.e., for symmetrical couplings. This expression has to be compared to $  \langle\widehat N\rangle_\text{series}=1-(2/\pi)\arctan(2\varepsilon_1/\Gamma)$ obtained from Eq.~(\ref{N_series}) when $\varepsilon_1=\varepsilon_2$. In the limit of large negative $\varepsilon_1$ compared to $\Gamma$, it leads to $N_\text{series}\approx 2$ and $N_\text{parallel}\approx 1$, explaining the reduction of $N$ by a factor two along the first diagonal (see Fig.~\ref{figure_N}). Physically, it corresponds to the decoupling of one of the two eigenstates of the DQD, the bonding eigenstate, from the leads.

\subsection{ML-DQD in parallel}\label{section_ML_DQD_parallel}

\begin{figure}[t]
\begin{center}
\hspace*{-0.3cm}\includegraphics[scale=.35]{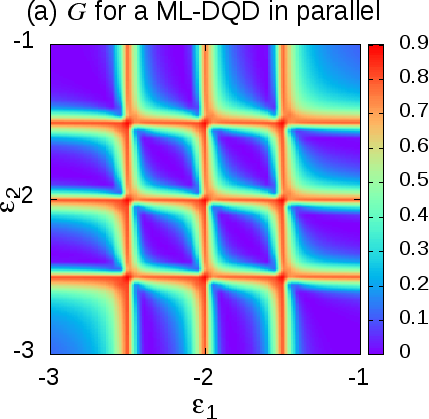}
\includegraphics[scale=.35]{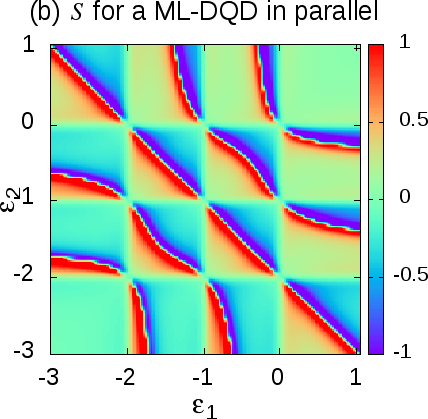}
\hspace*{-0.3cm}\includegraphics[scale=.35]{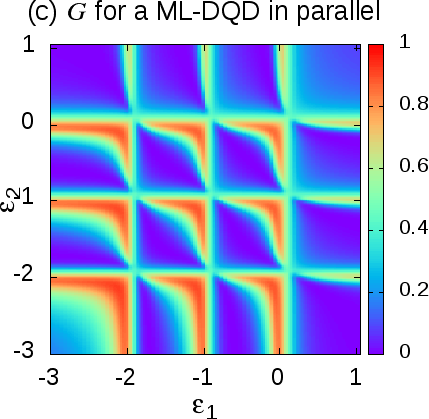}
\includegraphics[scale=.35]{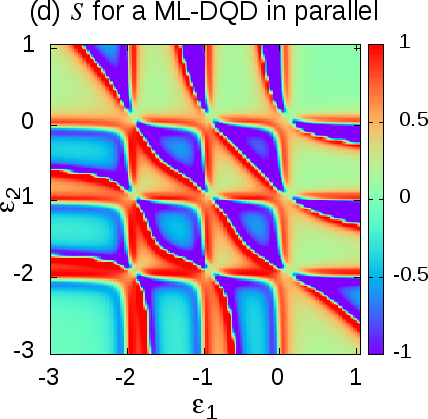}
\hspace*{-0.3cm}\includegraphics[scale=.35]{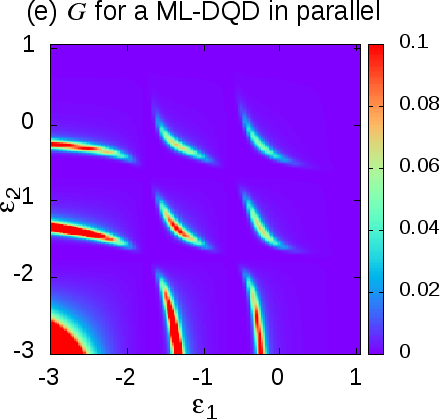}
\includegraphics[scale=.35]{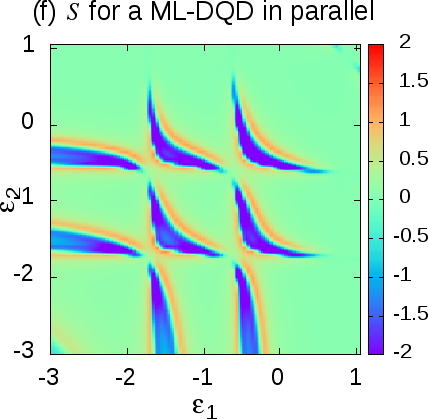}
\end{center}
\caption{Color-scale plots of $G$ and $S$ for a ML-DQD connected in parallel as a function of $\varepsilon_1$ and $\varepsilon_2$ at $\mu_{L,R}=0$, $k_BT=0.01$ for: (a), (b)~weak interdot coupling ($\m{V}_{12}=0.01$ and  $\Gamma_{\alpha,ij}=0.1$); (c), (d)~intermediate interdot coupling  ($\m{V}_{12}=\Gamma_{\alpha,ij}=0.1$); and (e), (f)~strong interdot coupling ($\m{V}_{12}=0.5$, $\Gamma_{\alpha,ij}=0.01$).}
\label{figure_G_2}
\end{figure}

\begin{figure}[t]
\hspace*{-0.3cm}\includegraphics[scale=.35]{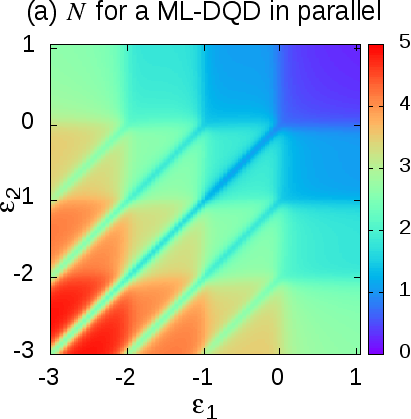}
\includegraphics[scale=.35]{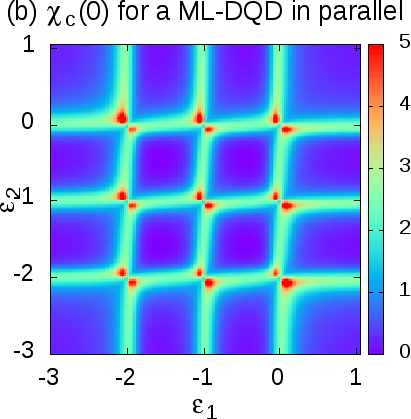}
\hspace*{-0.3cm}\includegraphics[scale=.35]{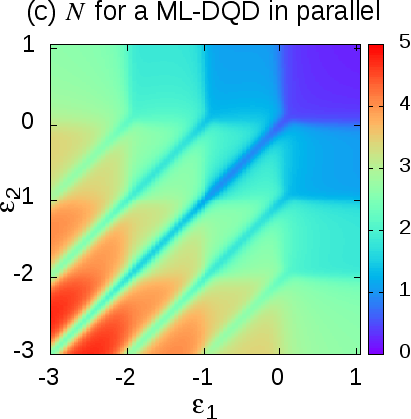}
\includegraphics[scale=.35]{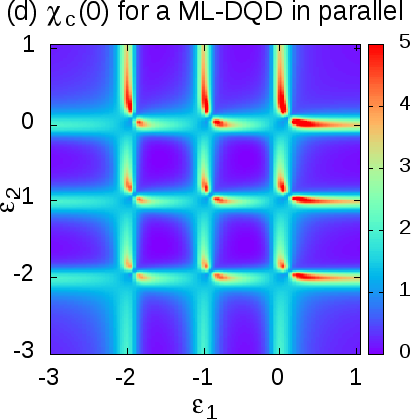}
\hspace*{-0.3cm}\includegraphics[scale=.35]{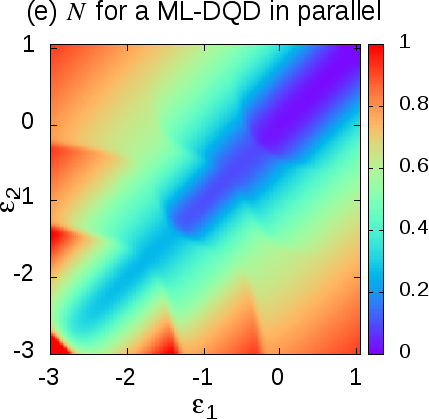}
\includegraphics[scale=.35]{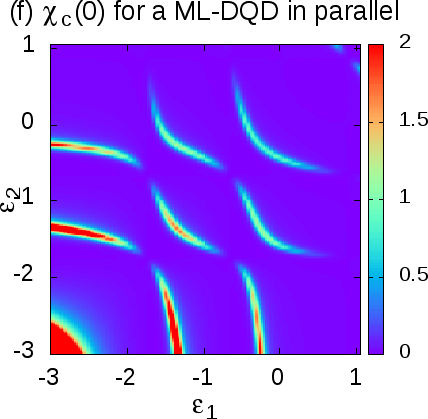}
\caption{Color-scale plots of $N$ and $\chi_c(0)$ for a ML-DQD connected in parallel as a function of $\varepsilon_1$ and $\varepsilon_2$.  The parameters are the same as in Fig.~\ref{figure_G_2}.}
\label{figure_S_2}
\end{figure}

We examine the case of a ML-DQQ in parallel schematically represented in Fig.~\ref{fig2}(d) taking as an example the case of three energy levels of energies as in the case of the ML-DQD in series. The results obtained for the color-scale plots of $G$, $S$ and $N$, $\chi_c(0)$ as a function of $\varepsilon_1$ and $\varepsilon_2$ are reported in Figs.~\ref{figure_G_2} and \ref{figure_S_2}, respectively, in the different interdot coupling regimes.

In the weak interdot coupling regime, i.e., for $|\m{V}_{12}| \ll \Gamma_{\alpha,ij}$, the color-scale plot for $G$ reported in Fig.~\ref{figure_G_2}(a) is strongly modified compared to the configuration in series, showing conductance ridges along the lines of a slightly distorted square lattice. The color-scale plot of $\chi_c(0)$ shown in Fig.~\ref{figure_S_2}(b) looks like that of $G$, with maxima along the lines of a lattice, with the presence of additional pairs of localized peaks around the nodes of the lattice. The results for $N$ reported in Fig.~\ref{figure_S_2}(a) reveal the charge stability diagram with the presence of $4\times 4=16$ domains as for the configuration in series. However one notices an important difference which is provided by the presence of elongated tips along the first diagonal as well as along the secondary first diagonals, with a dip of $N$ along them, similarly to what is observed in the case of the SL-DQD in parallel. Finally the color-scale plot of $S$ displayed in Fig.~\ref{figure_G_2}(b) shows that $S$ changes sign several times in the plane $(\varepsilon_1,\varepsilon_2)$.

In the intermediate interdot regime, i.e., for $|\m{V}_{12}|$ of the order of $\Gamma_{\alpha,ij}$, the results for the different color-scale plots of $G$, $S$, $N$ and $\chi_c(0)$ are strongly reminiscent of the results reported in Fig.~\ref{figure_SL_parallel} obtained in the case of the SL- DQD in parallel with the same choice of parameters.  The reported pattern corresponds to the duplication of the pattern observed in the single-level case, in each cell of a square lattice. Here again the presence of the elongated tips are clearly visible in the charge stability diagram revealed by the color-scale plot of $N$ displayed in Fig.~\ref{figure_S_2}(d).

In the strong interdot coupling regime, i.e., for $|\m{V}_{12}| \gg \Gamma_{\alpha,ij}$, the square lattice structure visible in the previous figures has disappeared, giving place to an oblique structure in the direction of the second diagonal as can be seen in Figs.~\ref{figure_G_2}(e) and \ref{figure_G_2}(f), and \ref{figure_S_2}(e) and \ref{figure_S_2}(f) for $G$, $S$, $N$, and $\chi_c(0)$ respectively. In addition, the extremities of the lines of maxima for $G$, $S$, and $\chi_c(0)$ disappear in the top-right part of the plane $(\varepsilon_1,\varepsilon_2)$ compared to the case of ML-DQD in series. The color-scale plot of $N$ shows the presence of a predominant tip along the first diagonal.

\section{Conclusion}\label{section_conclusion}

{  We have studied a noninteracting DQD system regardless of its geometry, either in series or in parallel, and analyzed} its electrical and thermoelectrical properties. In the case of single-level dots, the expressions for the nonequilibrium Green functions and electrical current are derived exactly. In the case of multilevel dots, an analytical calculation is performed assuming that the hopping integrals between the two dots and between the dots and the leads are independent of energy. The whole set of results apply to any temperatures, bias/gate voltages, and coupling strengths. The numerical results for the linear electrical conductance, the zero-frequency charge susceptibility, the Seebeck coefficient, and the dot occupancy are discussed in the light of previous works. In particular, the obtained results for $G$ and $\chi_c(0)$ show that with increasing interdot coupling, the system gradually changes from a regime where the two dots are almost decoupled to a regime where pairs of triple points have separated, until a regime where the two dots merge into a single one, in qualitative agreement with experiments, in particular in the case of a ML-DQD system in series. One observes that for a SL-DQD in series, the Seebeck coefficient undergoes three successive sign changes with increasing dot gates, again in good agreement with experiments. The cancellation of $S$ arises when the average energy of the charge carriers cancels, meaning that the system reaches an electron-hole symmetry situation. A level anticrossing effect resulting from finite interdot coupling manifests itself in the charge stability diagram with boundary lines separating the domains of different dot occupancies occurring when the energy levels of the bonding and antibonding states are aligned with the chemical potentials in the leads. In a striking way, we find a considerable reduction of the total dot occupancy in the case in parallel compared to the case in series, when the energy levels in each of the two dots are equal. We interpret this reduction as an effect of interferences produced by the presence of two transmission electronic paths in the parallel geometry and by the fact that the bonding eigenstate becomes disconnected from the leads. The approach developed in this paper can be directly used to study DQD systems driven out-of-equilibrium by applying either a finite bias voltage or a temperature gradient between the two leads, and/or in the presence of asymmetric couplings. The determination of the noise spectrum and finite-frequency charge susceptibility in the noninteracting DQD system is made in  Ref.~\onlinecite{Crepieux2020} following the theoretical approach developed in Refs.~\onlinecite{Zamoum2016,Crepieux2018}. A direct and essential extension of this work consists of taking into account the spin degrees of freedom and the Coulomb interactions in the dots.\\

{\it Acknowledgments} -- The authors would like to thank H. Baranger, S. Barraud, A. Crippa, A.-M. Dar\'e, S. De Franceschi, V. Kashcheyevs, R. Maurand, C. Mora, Y.-M. Niquet, S. Sahoo, M. Sanquer and R. Whitney for valuable discussions. They acknowledge the CEA-Programme Transverse Nanosciences and CEA-Eurotalents for financial support.


\appendix


{ 

\section{Electrical current in a DQD}\label{appendix_electrical_current}

\subsection{Equations of motion}\label{section_eom}

The Green functions in the dots are obtained by using the equation of motion approach. We adopt the Zubarev notation \cite{Zubarev1960} along which the retarded Green function in energy $G_{ \widehat A, \widehat B}^{r}(\varepsilon)$ associated with operators $ \widehat A$ and $ \widehat B$ is denoted by $\langle\langle  \widehat A; \widehat B\rangle\rangle$, with  $G_{ \widehat A, \widehat B}^{r}(\varepsilon)$ the Fourier transform of the retarded Green function in time, $G_{ \widehat A, \widehat B}^{r}(t,t')=-\Theta(t-t')\langle \{ \widehat A, \widehat B \} \rangle$, $\Theta$ being the Heaviside step function. By using this notation, the equation of motion for $ \langle\langle \widehat A; \widehat B \rangle\rangle$ writes
\begin{eqnarray}\label{eqom}
\varepsilon \langle\langle  \widehat A; \widehat B \rangle\rangle  &=& \langle \{ \widehat A, \widehat B \} \rangle +  \langle\langle [ \widehat A,\m{\widehat H}]; \widehat B \rangle\rangle
\end{eqnarray}
where $[ \widehat A,\m{\widehat H}]$ is the commutator between the operator $ \widehat A$ and the hamiltonian $\m{\widehat H}$ of the DQD system, and $\{ \widehat A, \widehat B\}$ is the anticommutator between the operators $ \widehat A$ and $ \widehat B$. Applying Eq.~(\ref{eqom}) to the various operators in the dots and the leads, one gets
\begin{eqnarray}
\varepsilon \langle\langle  \widehat d_{in}; \widehat d_{jm}^\dg\rangle\rangle&=&\delta_{ij}\delta_{nm} + \langle\langle \big[ \widehat d_{in},  \m{\widehat H}\big];  \widehat d_{jm}^\dg \rangle\rangle\\
\varepsilon \langle\langle  \widehat c_{\alpha k}; \widehat d_{jm}^\dg\rangle\rangle&=& \langle\langle \big[ \widehat c_{\alpha k},  \m{\widehat H}\big];  \widehat d_{jm}^\dg \rangle\rangle
\end{eqnarray}
since $ \{ \widehat d_{in}, \widehat d_{jm}^\dg\}=\delta_{ij}\delta_{nm}$ and $ \{ \widehat c_{\alpha k}, \widehat d_{jm}^\dg\}$=0. The calculation of the nonvanishing commutators between the operators $\widehat d_{in}$ and $\widehat c_{\alpha k}$ and the various terms of $\m{\widehat H}$ leads to
\begin{eqnarray}
&&\big[ \widehat d_{in}, \m{\widehat H}_\text{dots}\big] = \varepsilon_{in}  \widehat d_{in}+\sum_{n' \in \overline{i}} \m{V}^*_{in,\overline{i} n'} \widehat d_{\overline{i} n'}\\
&&\big[ \widehat d_{in},\m{\widehat H}_\text{hop}\big] =\sum_{\substack{\alpha =L,R\\k \in \alpha}} V_{in,\alpha k}^* \widehat c_{\alpha k}\\
&&\big[  \widehat c_{\alpha k}, \m{\widehat H}_\text{leads}\big] = \varepsilon_{\alpha k}  \widehat c_{\alpha k}\\
&&\big[ \widehat c_{\alpha k},\m{\widehat H}_\text{hop}\big] =\sum_{\substack{i=1,2\\n \in i}} V_{in,\alpha k} \widehat d_{in}
\end{eqnarray}
where $\overline{i}=2$ when $i=1$ and $\overline{i}=1$ when $i=2$. By collecting these contributions together and defining the retarded Green functions in the dots $G^{r}_{in,jm}(\varepsilon)=\langle\langle\widehat d_{in}; \widehat d_{jm}^\dg\rangle\rangle$, one gets the set of coupled equations
\begin{eqnarray}\label{eq1}
G^{r}_{in,jm}(\varepsilon) &=& \delta_{ij}\delta_{nm}g^{r}_{in}(\varepsilon) +  g^{r}_{in}(\varepsilon)\sum_{n' \in \overline{i}}\m{V}^*_{in,\overline{i} n'}G^{r}_{\overline{i}n',jm}(\varepsilon) \nonumber\\
&&+g^{r}_{in}(\varepsilon)\sum_{\substack{\alpha =L,R\\k \in \alpha}} V_{i n,\alpha k}^* G^{r}_{\alpha k,jm}(\varepsilon)\\\label{eq2}
G^{r}_{\alpha k,jm} (\varepsilon)&=&g^{r}_{\alpha k}(\varepsilon)\sum_{\substack{i'=1,2\\n' \in i'}} V_{i'n',\alpha k}G^{r}_{i'n',jm}(\varepsilon)
\end{eqnarray}
where $g_{in}^{r}(\varepsilon) =  (\varepsilon - \varepsilon_{in} + i0^+)^{-1}$ and $g_{\alpha k}^{r}(\varepsilon) =  (\varepsilon - \varepsilon_{\alpha k} + i0^+)^{-1}$ are the retarded Green functions in the disconnected dot $i$ and lead $\alpha$, respectively. By inserting Eq.~(\ref{eq2}) into Eq.~(\ref{eq1}), one finally obtains a Dyson-like equation
 \begin{eqnarray}\label{dyson}
&&G^{r}_{in,jm}(\varepsilon) = \delta_{ij}\delta_{nm}g^{r}_{in}(\varepsilon) +  \sum_{n' \in \overline{i}}g^{r}_{in}(\varepsilon)\m{V}^*_{in,\overline{i} n'}G^{r}_{\overline{i}n',jm}(\varepsilon) \nonumber\\
&&+\sum_{\substack{\alpha =L,R\\k \in \alpha}}\sum_{\substack{i'=1,2\\n' \in i'}}g^{r}_{in}(\varepsilon)V_{i n,\alpha k}^* g^{r}_{\alpha k}(\varepsilon)V_{i'n',\alpha k}G^{r}_{i'n',jm}(\varepsilon)
\end{eqnarray}
leading to a set of $4(N_\varepsilon\times M_\varepsilon)$ coupled linear equations. We want to underline that the presence of the term $ \sum_{n' \in \overline{i}}g^{r}_{in}(\varepsilon)\m{V}^*_{in,\overline{i} n'}G^{r}_{\overline{i}n',jm}(\varepsilon)$ in this equation is directly related to the fact that one has here two coupled dots.

In the next two sections, a distinction is made between the SL-DQD case for which an exact solution of  Eq.~(\ref{dyson}) can be derived and the ML-DQD case for which an approximate solution is given. This latter solution is obtained by making the assumption that the hopping integrals entering in the hamiltonian is independent both of the energy levels $\varepsilon_{in}$ in the dots and of the $k$-state in the leads.

\subsection{Exact result for a SL-DQD}\label{section_singlelevel}

When each of the two dots contains a single energy level, denoted as $\varepsilon_1$ and $\varepsilon_2$, the indices $n$, $n'$ and $m$ are absent and then Eq.~(\ref{dyson}) reduces to
 \begin{eqnarray}\label{dyson_single}
{G}^{r}_{ij}(\varepsilon)& =& \delta_{ij}{g}^{r}_{i}(\varepsilon) + { g}^{r}_{i}(\varepsilon)\m{V}^*_{i\overline{i} }{G}^{r}_{\overline{i}j} (\varepsilon)\nonumber\\
&&+\sum_{i'=1,2}{g}^{r}_{i}(\varepsilon){{\mathbb{\Sigma}}}_{\text{hop},ii'}^{r}(\varepsilon){ G}^{r}_{i'j}(\varepsilon)
\end{eqnarray}
where $g_{i}^{r}(\varepsilon) =  (\varepsilon - \varepsilon_{i} + i0^+)^{-1}$ is the retarded Green function of the disconnected single-level dot $i$, and ${{\mathbb{\Sigma}}}_{\text{hop},ii'}^{r}(\varepsilon)$ are the elements of the $2\times 2$ hopping self-energy matrix $\doubleunderline{{\mathbb{\Sigma}}}^{r}_\text{\;hop}(\varepsilon)$ in the $\{|1\rangle,|2\rangle\}$-basis
\begin{eqnarray}\label{def_sigma}
 \doubleunderline{{\mathbb{\Sigma}}}^{r}_\text{\;hop}(\varepsilon)= \sum_{\alpha=L,R}\sum_{k\in\alpha}g_{\alpha k}^{r}(\varepsilon)
\left(
\begin{array}{cc}
|V_{1,\alpha k}|^2 & V^*_{1,\alpha k}V_{2,\alpha k}\\
 V_{1,\alpha k}V^*_{2,\alpha k} & |V_{2,\alpha k}|^2
\end{array}
\right)\nonumber\\
\end{eqnarray}
In a matrix notation, the Dyson equation for $\doubleunderline{ G}^{r}(\varepsilon)$ writes
\begin{eqnarray}\label{dyson_matrix}
 \doubleunderline{ G}^{r}(\varepsilon)=\doubleunderline{g}^{r}(\varepsilon)+\doubleunderline{g}^{r}(\varepsilon)\doubleunderline{\mathbb{\Sigma}}^{r}(\varepsilon)\doubleunderline{ G}^{r}(\varepsilon)
\end{eqnarray}
where $\doubleunderline{g}^{r}(\varepsilon)$ is the retarded Green function matrix of the disconnected dots
\begin{eqnarray}
\doubleunderline{g}^{r}(\varepsilon)=\left(
\begin{array}{cc}
g_{1}^{r}(\varepsilon)&0\\
0&g_{2}^{r}(\varepsilon)
\end{array}
\right)
\end{eqnarray}
and $\doubleunderline{\mathbb{\Sigma}}^{r}(\varepsilon)$ is the retarded self-energy matrix given by
\begin{eqnarray}\label{def_sigma_r}
 \doubleunderline{\mathbb{\Sigma}}^{r}(\varepsilon)=\left(
\begin{array}{cc}
{\mathbb{\Sigma}}^r_{\text{hop},11}(\varepsilon) & {\mathbb{\Sigma}}^r_{\text{hop},12}(\varepsilon)+\m{V}^*_{12}\\
{\mathbb{\Sigma}}^r_{\text{hop},21}(\varepsilon)+\m{V}^*_{21} & {\mathbb{\Sigma}}^r_{\text{hop},22}(\varepsilon)
\end{array}
\right)
\end{eqnarray}
Equation~(\ref{dyson_matrix}) can be solved exactly. The calculation presented hereafter applies to the case of a SL-DQD but can be generalized to the case of a ML-DQD within the limit of the approximations made in this paper. The explicit dependences with  energy $\varepsilon$ of the Green functions and of the self-energies are omitted in the next equations in order to lighten the notations. Starting from Eq.~(\ref{dyson_single}) one writes the four equations of motion followed by the elements of the matrix $\doubleunderline{  G}^r$
\begin{eqnarray}
 {  G}^{r}_{11}& =&{  g}^{r}_{1} + {  g}^{r}_{1}\m{V}^*_{12}{  G}^{r}_{21} 
+{  g}^{r}_{1}({{\mathbb{\Sigma}}}_{\text{hop},11}^{r}{  G}^{r}_{11}+{{\mathbb{\Sigma}}}_{\text{hop},12}^{r}{  G}^{r}_{21})\nonumber\\
 {  G}^{r}_{12}& =& {  g}^{r}_{1}\m{V}^*_{12}{  G}^{r}_{22} 
+{  g}^{r}_{1}({{\mathbb{\Sigma}}}_{\text{hop},11}^{r}{  G}^{r}_{12}+{{\mathbb{\Sigma}}}_{\text{hop},12}^{r}{  G}^{r}_{22})\nonumber\\
 {  G}^{r}_{21}& =& {  g}^{r}_{2}\m{V}^*_{21}{  G}^{r}_{11} 
+{  g}^{r}_{2}({{\mathbb{\Sigma}}}_{\text{hop},21}^{r}{  G}^{r}_{11}+{{\mathbb{\Sigma}}}_{\text{hop},22}^{r}{  G}^{r}_{21})\nonumber\\
 {  G}^{r}_{22}& =&{  g}^{r}_{2} + {  g}^{r}_{2}\m{V}^*_{21}{  G}^{r}_{12} 
+{  g}^{r}_{2}({{\mathbb{\Sigma}}}_{\text{hop},21}^{r}{  G}^{r}_{12}+{{\mathbb{\Sigma}}}_{\text{hop},22}^{r}{  G}^{r}_{22})\nonumber
\end{eqnarray}
This set of linear equations can be rewritten in the following matrix form $\doubleunderline{  M}^r(\varepsilon)\doubleunderline{  G}^r(\varepsilon)=\doubleunderline{  g}^r(\varepsilon)$, where
\begin{eqnarray}\label{def_M}
\doubleunderline{  M}^r=
\left(\begin{array}{cc}
  1-{  g}^{r}_{1}{{\mathbb{\Sigma}}}_{11}^{r} & -{  g}^{r}_{1}\mathbb{\Sigma}^r_{12}\\
- {  g}^{r}_{2}\mathbb{\Sigma}^r_{21} &  1-{  g}^{r}_{2}{{\mathbb{\Sigma}}}_{22}^{r}
 \end{array}\right)
\end{eqnarray}
with $\mathbb{\Sigma}^r_{ii}(\varepsilon)={{\mathbb{\Sigma}}}_{\text{hop},ii}^{r}(\varepsilon)$,  $\mathbb{\Sigma}^r_{i\overline{i}}(\varepsilon)={{\mathbb{\Sigma}}}_{\text{hop},i\overline{i}}^{r}(\varepsilon)+\m{V}^*_{i\overline{i}}$ and,
\begin{eqnarray}
\doubleunderline{  G}^r(\varepsilon)&=&
 \left(\begin{array}{cc}
 {  G}^{r}_{11}(\varepsilon) &  {  G}^{r}_{12}(\varepsilon)\\
 {  G}^{r}_{21}(\varepsilon) &  {  G}^{r}_{22}(\varepsilon)
 \end{array}\right)\\
\doubleunderline{  g}^r(\varepsilon)
&=&\left(\begin{array}{cc}
  {  g}^{r}_{1}(\varepsilon) & 0\\
0 & {  g}^{r}_{2}(\varepsilon)
 \end{array}\right)
\end{eqnarray}
The solution of this matrix equation is given by $\doubleunderline{  G}^r(\varepsilon)=(\doubleunderline{  M}^r(\varepsilon))^{-1}\doubleunderline{  g}^r(\varepsilon)$, one obtains the following expression for the $2\times 2$ Green function matrix $\doubleunderline{G}^{r}(\varepsilon)$
\begin{eqnarray}\label{FG_single}
\doubleunderline{ G}^{r}(\varepsilon) =
\frac{1}{D^{r}(\varepsilon)}\left(
\begin{array}{cc}
\widetilde{ g}_1^{\;r}(\varepsilon) & \widetilde{  g}_1^{\;r}(\varepsilon)\mathbb{\Sigma}_{12}^{r}(\varepsilon)\widetilde{  g}_2^{\;r}(\varepsilon)\\
\widetilde{  g}_2^{\;r}(\varepsilon)\mathbb{\Sigma}_{21}^{r}(\varepsilon)\widetilde{  g}_1^{\;r}(\varepsilon) & \widetilde{  g}_2^{\;r}(\varepsilon)
\end{array}
\right)\nonumber\\
\end{eqnarray}
where ${\widetilde{g}_{i}^{\;r}}(\varepsilon)$ is defined by ${\widetilde{g}_{i}^{\;r}}(\varepsilon)={g}_{i}^{r} (\varepsilon)/(1 - {{{\mathbb{\Sigma}}}_{\text{hop},ii}^{r}}(\varepsilon) {g}_{i}^{r}(\varepsilon))$,  $D^{r}(\varepsilon)$ is given by $D^{r}(\varepsilon) = 1-\widetilde{g}_{1}^{\;r}(\varepsilon)\mathbb{\Sigma}_{12}^{r}(\varepsilon)\widetilde{g}_{2}^{\;r}(\varepsilon)\mathbb{\Sigma}_{21}^{r}(\varepsilon)$. Equation~(\ref{FG_single}) gives the exact expression of the retarded Green function $\doubleunderline{G}^{r}(\varepsilon)$ in a SL-DQD.  It holds as well as for serial as for parallel geometries of the DQD system. We want to underline that for a DQD connected in series, the self-energy  $ \doubleunderline{{\mathbb{\Sigma}}}^{r}_\text{hop}(\varepsilon)$ defined in Eq.~(\ref{def_sigma}) becomes a diagonal matrix since the product $V^*_{1,\alpha k}V_{2,\alpha k}$ is equal to zero whatever the index $\alpha$ is. Therefore, the off-diagonal elements of the total self-energy matrix $ \doubleunderline{\mathbb{\Sigma}}^{r}$ defined in Eq.~(\ref{def_sigma_r}) reduces to $\m{V}^*_{12}$ and $\m{V}^*_{21}$ in that case.

\subsection{Generalization to a ML-DQD}\label{section_multilevel}

In realistic systems, the dots constituting the DQD system contain several energy levels, as for example in Ge/Si heterostructure nanowire-based DQDs \cite{Hu2007} or in graphene-based DQDs \cite{Li2013}. In that situation one would have to perform a numerical calculation to determine the solutions of Eq.~(\ref{dyson}). However, when the hopping integrals $\m{V}_{in,\overline{i}m}$ and $V_{in,\alpha k}$ do not depend on the indices $n$ and $m$ and on the state $k$ (and in that case, we use the notations $\m{V}_{i\overline{i}}$ and $V_{i\alpha}$), the calculation remains analytical. Within this assumption and by performing a double sum over the $n$ and $m$ indices, Eq.~(\ref{dyson}) becomes
 \begin{eqnarray}\label{dyson_sum}
{\bf G}^{r}_{ij}(\varepsilon)& =& \delta_{ij}{\bf g}^{r}_{i}(\varepsilon) + {\bf g}^{r}_{i}(\varepsilon)\m{V}^*_{i\overline{i} }{\bf G}^{r}_{\overline{i}j} (\varepsilon)\nonumber\\
&&+\sum_{i'=1,2}{\bf g}^{r}_{i}(\varepsilon){\mathbb{\Sigma}}_{\text{hop},ii'}^{r}(\varepsilon){\bf G}^{r}_{i'j}(\varepsilon)
\end{eqnarray}
where
\begin{eqnarray}\label{Gtot}
&&{\bf G}^{r}_{ij}(\varepsilon)=\sum_{n\in i, m\in j}G^{r}_{in,jm}(\varepsilon)\\
&&{\bf g}_i^{r}(\varepsilon)=\sum_{n\in i}g_{in}^{r}(\varepsilon)\\
&&{\mathbb{\Sigma}}_{\text{hop},ij}^{r}(\varepsilon)=\sum_{\alpha=L,R}\sum_{k\in \alpha}V_{i\alpha}^*g^r_{\alpha k}(\varepsilon)V_{j\alpha}
\end{eqnarray}
In a matrix form, Eq.~(\ref{dyson_sum}) reads as
\begin{eqnarray}\label{dyson_matrix_ML}
 \doubleunderline{\bf G}^{r}(\varepsilon)=\doubleunderline{\bf g}^{r}(\varepsilon)+\doubleunderline{\bf g}^{r}(\varepsilon)\doubleunderline{\mathbb{\Sigma}}^{r}(\varepsilon)\doubleunderline{\bf G}^{r}(\varepsilon)
\end{eqnarray}
The solutions of Eq.~(\ref{dyson_matrix_ML}) can be obtained analytically since it is a set of four linear equations. In matrix notation we obtain a $2\times 2$ matrix $\doubleunderline{\bf G}^{r}(\varepsilon)$, the elements of which correspond to ${\bf G}^{r}_{ij}(\varepsilon)$,
\begin{eqnarray}\label{FG}
\doubleunderline{\bf G}^{r}(\varepsilon) =
\frac{1}{D^{r}(\varepsilon)}\left(
\begin{array}{cc}
\widetilde{\bf g}_1^{r}(\varepsilon) & \widetilde{\bf g}_1^{r}(\varepsilon)\mathbb{\Sigma}_{12}^{r}(\varepsilon)\widetilde{\bf g}_2^{r}(\varepsilon)\\
\widetilde{\bf g}_2^{r}(\varepsilon)\mathbb{\Sigma}_{21}^{r}(\varepsilon)\widetilde{\bf g}_1^{r}(\varepsilon) & \widetilde{\bf g}_2^{r}(\varepsilon)
\end{array}
\right)\nonumber\\
\end{eqnarray}
with $D^{r}(\varepsilon) = 1-\widetilde{\bf g}_{1}^{r}(\varepsilon)\mathbb{\Sigma}_{12}^{r}(\varepsilon)\widetilde{\bf g}_{2}^{r}(\varepsilon)\mathbb{\Sigma}_{21}^{r}(\varepsilon)$,  ${\widetilde{\bf g}_{i}^{r}}(\varepsilon) ={\bf g}_{i}^{r} (\varepsilon)/(1 - {{{\mathbb{\Sigma}}}_{\text{hop},ii}^{r}}(\varepsilon){\bf g}_{i}^{r}(\varepsilon))$ and $\mathbb{\Sigma}_{ij}^{r}(\varepsilon)= {{\mathbb{\Sigma}}}_{\text{hop},ij}^r(\varepsilon)+\delta_{j\overline{i}}\m{V}^*_{i\overline{i}}$. Equation~(\ref{FG}) provides the expression of the retarded Green function $\doubleunderline{\bf G}^{r}(\varepsilon) $ in a ML-DQD within the assumption that the hopping integrals are independent of the energy levels. We remark that this result is similar to Eq.~(\ref{FG_single}) obtained for a SL-DQD, provided that $\doubleunderline{\bf G}^{r}(\varepsilon)$, $\widetilde{\bf g}_i^{\;r}(\varepsilon)$ are changed into $\doubleunderline{G}^{r}(\varepsilon)$, $\widetilde{g}_i^{r}(\varepsilon)$. The advanced Green function $\doubleunderline{\bf G}^{a}(\varepsilon)$ is obtained straightforwardly by replacing the superscript $r$ by the superscript $a$ in Eq.~(\ref{FG}) with $D^{a}(\varepsilon) = 1-\widetilde{\bf g}_{1}^{a}(\varepsilon)\mathbb{\Sigma}_{12}^{a}(\varepsilon)\widetilde{\bf g}_{2}^{a}(\varepsilon)\mathbb{\Sigma}_{21}^{a}(\varepsilon)$ where
${\widetilde{\bf g}_{i}^{a}}(\varepsilon) ={\bf g}_{i}^{a} (\varepsilon)/(1 -{{{\mathbb{\Sigma}}}_{\text{hop},ii}^{a}(\varepsilon) {\bf g}_{i}^{a} (\varepsilon)})$ and  $\mathbb{\Sigma}_{ij}^{a}(\varepsilon)= {{\mathbb{\Sigma}}}_{\text{hop},ij}^a(\varepsilon)+\delta_{j\overline{i}}\m{V}_{i\overline{i}}$.

To be able to describe the out-of-equilibrium properties of the DQD such as the electrical current and the electrical conductance, it is necessary to also determine the lesser and greater Green functions $\doubleunderline{\bf G}^{\lessgtr}(\varepsilon)$  for the DQD system as detailed in the next section.

\subsection{Lesser and greater Green functions}\label{appendix_keldysh_green_functions}

The lesser Green function matrix $\doubleunderline{{\bf G}}^{<}(\varepsilon)$ can be obtained by using the Langreth analytic continuation rules \cite{Langreth1991} on the Dyson equation for the contour ordered Keldysh Green functions obtained from Eq.~(\ref{dyson_matrix_ML}). One gets 
\begin{eqnarray}\label{dyson_less}
 &&\doubleunderline{\bf G}^{<}(\varepsilon)=\doubleunderline{\bf g}^{<}(\varepsilon)+\doubleunderline{\bf g}^{r}(\varepsilon)\doubleunderline{\mathbb{\Sigma}}^{r}(\varepsilon)\doubleunderline{\bf G}^{<}(\varepsilon)\nonumber\\
&&+\doubleunderline{\bf g}^{r}(\varepsilon)\doubleunderline{\mathbb{\Sigma}}^{<}(\varepsilon)\doubleunderline{\bf G}^{a}(\varepsilon)
+\doubleunderline{\bf g}^{<}(\varepsilon)\doubleunderline{\mathbb{\Sigma}}^{a}(\varepsilon)\doubleunderline{\bf G}^{a}(\varepsilon)
\end{eqnarray}
By performing successive iterations on $\doubleunderline{\bf G}^{<}(\varepsilon)$ in the r.h.s. of Eq.~(\ref{dyson_less}), one obtains
\begin{eqnarray}\label{dyson_lesser}
 &&\doubleunderline{\bf G}^{<}(\varepsilon)=\doubleunderline{\bf G}^{r}(\varepsilon)\doubleunderline{\mathbb{\Sigma}}^{<}(\varepsilon)\doubleunderline{\bf G}^{a}(\varepsilon)\nonumber\\
&&+\left[1+\doubleunderline{\bf G}^{r}(\varepsilon)\doubleunderline{\mathbb{\Sigma}}^{r}(\varepsilon)\right]\doubleunderline{\bf g}^{<}(\varepsilon)
\left[1+\doubleunderline{\mathbb{\Sigma}}^{a}(\varepsilon)\doubleunderline{\bf G}^{a}(\varepsilon)\right]
\end{eqnarray}
The second term in the r.h.s of Eq.~(\ref{dyson_lesser}) vanishes since it can be put in the form
\begin{eqnarray}
 \left[1+\doubleunderline{\bf G}^{r}(\varepsilon)\doubleunderline{\mathbb{\Sigma}}^{r}(\varepsilon)\right]\doubleunderline{\bf g}^{<}(\varepsilon)
\left[1+\doubleunderline{\mathbb{\Sigma}}^{a}(\varepsilon)\doubleunderline{\bf G}^{a}(\varepsilon)\right]\nonumber\\
=\doubleunderline{\bf G}^{r}(\varepsilon)(\doubleunderline{\bf g}^{r}(\varepsilon))^{-1}\doubleunderline{\bf g}^{<}(\varepsilon)(\doubleunderline{\bf g}^{a}(\varepsilon))^{-1}\doubleunderline{\bf G}^{a}(\varepsilon)
\end{eqnarray}
with $(\doubleunderline{\bf g}^{r}(\varepsilon))^{-1}\doubleunderline{\bf g}^{<}(\varepsilon)(\doubleunderline{\bf g}^{a}(\varepsilon))^{-1}=\doubleunderline{0}$, stemming from the fact that $\doubleunderline{\bf g}(\varepsilon)$ is the Green function for the disconnected noninteracting DQD system\cite{Jauho1994}. Therefore, and generalizing it to the greater Green functions $\doubleunderline{\bf G}^{>}(\varepsilon)$, one has
\begin{eqnarray}\label{FGK}
 \doubleunderline{\bf G}^{\lessgtr}(\varepsilon)=\doubleunderline{\bf G}^{r}(\varepsilon)\doubleunderline{{\mathbb{\Sigma}}}^{\lessgtr}(\varepsilon)\doubleunderline{\bf G}^{a}(\varepsilon)
\end{eqnarray}
where the lesser and greater self-energies $ \doubleunderline{{\mathbb{\Sigma}}}^{\lessgtr}(\varepsilon)$ are given by
\begin{eqnarray}
 \doubleunderline{{\mathbb{\Sigma}}}^{\lessgtr}(\varepsilon)= \sum_{\alpha=L,R}\sum_{k\in\alpha}g_{\alpha k}^{\lessgtr}(\varepsilon)
\left(
\begin{array}{cc}
|V_{1\alpha}|^2 & V^*_{1\alpha}V_{2\alpha}\\
 V_{1\alpha}V^*_{2\alpha} & |V_{2\alpha}|^2
\end{array}
\right)
\end{eqnarray}
Whereas $\doubleunderline{{\mathbb{\Sigma}}}^{r,a}(\varepsilon)$ differs from $ \doubleunderline{{\mathbb{\Sigma}}}^{r,a}_\text{\;hop}(\varepsilon)$ by the off-diagonal terms $\m{V}_{12}^*$ and $\m{V}_{21}^*$, $\doubleunderline{{\mathbb{\Sigma}}}^{\lessgtr}(\varepsilon)$ coincides with $ \doubleunderline{{\mathbb{\Sigma}}}^{\lessgtr}_\text{\;hop}(\varepsilon)$; thus one can indifferently use one or the other in any expression where these quantities appear.

The result expressed in Eq.~(\ref{FGK}) is remarkably simple. It indicates that the information about the inner details of the DQD system is entirely coded in the retarded/advanced Green functions $\doubleunderline{\bf G}^{r,a}(\varepsilon)$. We underline that the calculation of $ \doubleunderline{\bf G}^{\lessgtr}(\varepsilon)$ is made here for a ML-DQD system. However one can immediately deduce the lesser and greater Green functions $ \doubleunderline{G}^{\lessgtr}(\varepsilon)$ for the SL-DQD system by simply changing $\doubleunderline{\bf G}^{r,a}(\varepsilon)$ into $\doubleunderline{G}^{r,a}(\varepsilon)$ in Eq.~(\ref{FGK}). This also applies to the next section.

\subsection{Explicit expression for the electrical current}\label{appendix_explicit_expression_for_current}

The current operator from the lead $\alpha$ is defined as $\widehat I_\alpha(t)=-e d\widehat N_\alpha(t)/dt$ with $\widehat N_\alpha(t)=\sum_{k\in\alpha}\widehat c^\dg_{\alpha k}(t)\widehat c_{\alpha k}(t)$. In the steady state the derivative with respect to the time variable is given by $d\widehat N_\alpha(t)/dt=[\widehat N_\alpha(t),\m{\widehat  H}]/i\hbar$\cite{Caroli1971}. Thus the average current writes
\begin{eqnarray}
 I_\alpha=\langle \widehat I_\alpha \rangle=-\frac{e}{i\hbar}\sum_{k\in\alpha}\langle [\widehat c^\dg_{\alpha k}(t)\widehat c_{\alpha k}(t),\m{\widehat  H}]\rangle
\end{eqnarray}
The only term in $\m{\widehat  H}$ leading to a nonvanishing commutator with the product of operators $\widehat c^\dg_{\alpha k}(t)\widehat c_{\alpha k}(t)$ is $\m{\widehat H}_\text{hop}$. One gets
 \begin{eqnarray}\label{I_green}
 I_\alpha=\frac{e}{\hbar}\sum_{k\in\alpha}\sum_{\substack{i=1,2\\n \in i}}\left( V_{i\alpha} G^<_{in,\alpha k}(t,t)- V^*_{i\alpha} G^<_{\alpha k,in}(t,t)\right)\nonumber\\
\end{eqnarray}
where one has defined the lesser and greater Green functions $G^<_{in,\alpha k}(t,t')=i\langle\widehat c^\dg_{\alpha k}(t') \widehat d_{in}(t)\rangle$ and  $G^<_{\alpha k,in}(t,t')=i\langle \widehat d^\dg_{in}(t')\widehat c_{\alpha k}(t)\rangle$. Performing a Fourier transform, one gets
 \begin{eqnarray}
 I_\alpha=\frac{e}{h}\sum_{k\in\alpha}\sum_{\substack{i=1,2\\n \in i}}\int_{-\infty}^{\infty}\left( V_{i\alpha} G^<_{in,\alpha k}(\varepsilon)- V^*_{i\alpha} G^<_{\alpha k,in}(\varepsilon)\right)d\varepsilon\nonumber\\
\end{eqnarray}
In order to calculate the lesser and greater Green functions $G^<_{in,\alpha k}(\varepsilon)$ and $G^<_{\alpha k,in}(\varepsilon)$, one applies the Langreth analytic continuation rules\cite{Langreth1991}. From Eq.~(\ref{eq2}), one obtains
\begin{eqnarray}
G^{<}_{\alpha k,in}(\varepsilon) =\sum_{\substack{j=1,2\\m \in j}}V_{j\alpha}\Big(g^{r}_{\alpha k}(\varepsilon) G^{<}_{jm,in}(\varepsilon)\nonumber\\
+g^{<}_{\alpha k}(\varepsilon)G^{a}_{jm,in}(\varepsilon)\Big)
\end{eqnarray}
Similarly,
\begin{eqnarray}
G^{<}_{in,\alpha k}(\varepsilon) =\sum_{\substack{j=1,2\\m \in j}}V^*_{j\alpha}\Big(G^{<}_{in,jm}(\varepsilon)g^{a}_{\alpha k}(\varepsilon) \nonumber\\
+G^{r}_{in,jm}(\varepsilon)g^{<}_{\alpha k}(\varepsilon)\Big)
\end{eqnarray}
By inserting these expressions into Eq.~(\ref{I_green}), one gets
\begin{eqnarray}
 I_\alpha&=&\frac{e}{h}\sum_{k\in\alpha}\sum_{i=1,2}\sum_{j=1,2}\int_{-\infty}^{\infty}V_{i\alpha}V^*_{j\alpha}\nonumber\\
&&\times\Big(\big(g^a_{\alpha k}(\varepsilon)-g^r_{\alpha k}(\varepsilon)\big) {\bf G}^<_{ij}(\varepsilon)\nonumber\\
&&+g^<_{\alpha k}(\varepsilon)\big( {\bf G}^r_{ij}(\varepsilon)- {\bf G}^a_{ij}(\varepsilon)\big)\Big)d\varepsilon
\end{eqnarray}
where ${\bf G}^r_{ij}(\varepsilon)$ is the Green function summed over the indices $n$ and $m$ as defined in Eq.~(\ref{Gtot}). By using the general relationship $ \m{G}^r(\varepsilon)- \m{G}^a(\varepsilon)= \m{G}^>(\varepsilon)- \m{G}^<(\varepsilon)$ which holds for any Green function $\m{G}$, one obtains
\begin{eqnarray}\label{I_alpha}
 I_\alpha&=&\frac{e}{h}\sum_{k\in\alpha}\sum_{i=1,2}\sum_{j=1,2}\int_{-\infty}^{\infty}V_{i\alpha}V^*_{j\alpha}\nonumber\\
&&\times\Big(g^<_{\alpha k}(\varepsilon) {\bf G}^>_{ij}(\varepsilon)
-g^>_{\alpha k}(\varepsilon){\bf G}^<_{ij}(\varepsilon)\Big)d\varepsilon
\end{eqnarray}
Physically, this expression is interpreted as follows: the first contribution in $I_\alpha$ represents the current flowing from the $\alpha$ lead to the DQD since it is the product of the out-tunneling rate of the occupied state in the $\alpha$ lead, $\sum_{k\in\alpha}V_{i\alpha}V^*_{j\alpha}g^<_{\alpha k}$ which corresponds to the self-energy, and of the number of unoccupied states in the DQD, ${\bf G}^>_{ij}(\varepsilon)$, whereas the second contribution with the minus sign corresponds to the current flowing from the DQD to the lead $\alpha$. Equation~(\ref{I_alpha}) can be written thanks to Eq.~(\ref{FGK}) under the following form
 \begin{eqnarray}\label{I_inter}
 I_\alpha&=&\frac{e}{h}\int_{-\infty}^{\infty}
\text{Tr}\Big[\doubleunderline{{\mathbb{\Sigma}}}_{\;\alpha}^{<}(\varepsilon) \doubleunderline{\bf G}^{r}(\varepsilon)\doubleunderline{{\mathbb{\Sigma}}}^{>}(\varepsilon) \doubleunderline{\bf G}^{a}(\varepsilon)\nonumber\\
&&-\doubleunderline{{\mathbb{\Sigma}}}_{\;\alpha}^{>}(\varepsilon)\doubleunderline{\bf G}^{r}(\varepsilon)\doubleunderline{{\mathbb{\Sigma}}}^{<}(\varepsilon) \doubleunderline{\bf G}^{a}(\varepsilon)\Big]d\varepsilon
\end{eqnarray}
where the matrix elements of the self-energy $\doubleunderline{{\mathbb{\Sigma}}}_{\;\alpha}^{\lessgtr}(\varepsilon)$ are defined as $ {\Sigma}_{\alpha,ij}^{\lessgtr}(\varepsilon)=\sum_{k\in\alpha} V_{i\alpha}^* g_{\alpha k}^{\lessgtr}(\varepsilon)V_{j\alpha}$ and where $\text{Tr}[\;]$ denotes the trace of the matrix. In the limit of wide flat band for elecrons in the leads and energy-independent hopping integrals, one has
\begin{eqnarray}\label{sigma_less}
\doubleunderline{{\mathbb{\Sigma}}}_{\;\alpha}^{<}(\varepsilon)&=&i  f_\alpha(\varepsilon)\doubleunderline{\Gamma}_{\;\alpha}\\\label{sigma_great}
\doubleunderline{{\mathbb{\Sigma}}}_{\;\alpha}^{>}(\varepsilon)&=&-i (1-f_\alpha(\varepsilon))\doubleunderline{\Gamma}_{\;\alpha}
\end{eqnarray}
where the elements $\Gamma_{\alpha,ij}$ of the matrix $\doubleunderline{\Gamma}_{\;\alpha}$ are defined as $\Gamma_{\alpha,ij}=2\pi\rho_\alpha V^*_{i\alpha}V_{j\alpha}$ with $\rho_\alpha$, the density of states in the lead $\alpha$. We also have $\doubleunderline{{\mathbb{\Sigma}}}_{\;\alpha}^{r,a}(\varepsilon)=\mp i\doubleunderline{\Gamma}_{\;\alpha}/2$. By inserting Eqs.~(\ref{sigma_less}) and (\ref{sigma_great}) into Eq.~(\ref{I_inter}), one obtains
\begin{eqnarray}\label{I_final}
  I_\alpha=\frac{e}{h}\int_{-\infty}^{\infty}
&& \m{T}_{\alpha\overline{\alpha}}(\varepsilon)\big(f_\alpha(\varepsilon)-f_{\overline{\alpha}}(\varepsilon)\big)d\varepsilon
\end{eqnarray}
which corresponds to Eq.~(\ref{I_final_maintext}), where $ \m{T}_{\alpha\overline{\alpha}}(\varepsilon)$ is the transmission coefficient equal to
\begin{eqnarray}
 \m{T}_{\alpha\overline{\alpha}}(\varepsilon)=\text{Tr}\Big[\doubleunderline{\Gamma}_{\;\alpha}\;\doubleunderline{\bf G}^{r}(\varepsilon)\;\doubleunderline{\Gamma}_{\;\overline{\alpha}}\; \doubleunderline{\bf G}^{a}(\varepsilon)\Big]
\end{eqnarray}
with $\overline{\alpha}=R$ for $\alpha=L$ and $\overline{\alpha}=L$ for $\alpha=R$, and where $\doubleunderline{\Gamma}_{\;\alpha} $ is the dot-lead coupling matrix defined as
\begin{eqnarray}
\doubleunderline{\Gamma}_{\;\alpha} =2\pi\rho_\alpha\left(
\begin{array}{cc}
|V_{1\alpha}|^2 & V^*_{1\alpha}V_{2\alpha}\\
V_{1\alpha}V^*_{2\alpha} & |V_{2\alpha}|^2
\end{array}
\right)
\end{eqnarray}

}


\section{$\m{\widehat H}_\text{dots}$ eigenvalues}\label{appendix_Hdot}

In this appendix, we determine the eigenenergies and eigenvectors of the hamiltonian $\m{\widehat H}_\text{dots}$ of Eq.~(\ref{Hdot}) describing the DQD disconnected from the leads firstly for a SL-DQD and secondly for a ML-DQD with three levels of energy.

\subsection{SL-DQD}

\begin{figure}
\hspace*{-0.5cm}\includegraphics[scale=.5]{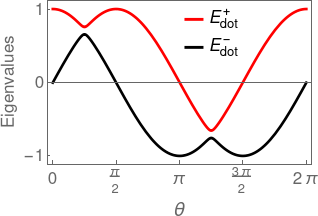}
\caption{Eigenenergies $E^+_\text{dots}$ (red curve) and $E^-_\text{dots}$ (black curve) at $\mathcal{V}_{12}=0.1$ as a function of the angle $\theta$ with $\varepsilon_1=\cos(\theta)$ and $\varepsilon_2=\sin(\theta)$. Level anticrossing effect arises at $\theta=\pi/4$ or $\theta=5\pi/4$, i.e. at $\varepsilon_1=\varepsilon_2$, with a distance between the red and black curves equal to $\Delta E=2|\mathcal{V}_{12}|=0.2$.}
\label{figure_AC}
\end{figure} 

For a SL-DQD, the hamiltonian $\m{\widehat H}_\text{dots}$ writes as a $2\times 2$ matrix in the basis $\{|1\rangle,|2\rangle\}$ of the states in the two dots 1 and 2
\begin{eqnarray}
 \m{\widehat H}_\text{dots}=
\left(
\begin{array}{cc}
\varepsilon_1 & \mathcal{V}_{12}^*\\
\mathcal{V}_{21}^*&\varepsilon_2
\end{array}
\right)
\end{eqnarray}
It can be diagonalized leading to the following eigenenergies $ E^+_\text{dots}$ and $ E^-_\text{dots}$
\begin{eqnarray}\label{eig_2}
 E^\pm_\text{dots}=\frac{\varepsilon_1+\varepsilon_2\pm\sqrt{(\varepsilon_1-\varepsilon_2)^2+4|\mathcal{V}_{12}|^2}}{2}
\end{eqnarray}
and eigenvectors $|+\rangle$ and $|-\rangle$ which correspond to the antibonding and bonding eigenstates of the SL-DQD
\begin{eqnarray}\label{def_EV1}
 |+\rangle&=&u^*|1\rangle+v^*|2\rangle\\\label{def_EV2}
|-\rangle&=&-v|1\rangle+u|2\rangle
\end{eqnarray}
with
\begin{eqnarray}\label{def_u}
 |u|^2&=&\frac{1}{2}\left(1+\frac{\varepsilon_1-\varepsilon_2}{E^+_\text{dots}-E^-_\text{dots}}\right)\\\label{def_v}
|v|^2&=&\frac{1}{2}\left(1-\frac{\varepsilon_1-\varepsilon_2}{E^+_\text{dots}-E^-_\text{dots}}\right)
\end{eqnarray}
and therefore $|uv|^2=|\m{V}_{12}|^2/(E^+_\text{dots}-E^-_\text{dots})^2$. It gives rise to a level anticrossing effect as soon as the interdot coupling $\mathcal{V}_{12}$ is finite. The anticrossing of the two levels $\varepsilon_1$ and $\varepsilon_2$ occurs in the vicinity of the first diagonal as shown in Fig.~\ref{figure_AC}. Indeed, from Eq.~(\ref{eig_2}), the difference between the two eigenenergies reads as $\Delta E=E^+_\text{dots}-E^-_\text{dots}=\sqrt{(\varepsilon_1-\varepsilon_2)^2+4|\mathcal{V}_{12}|^2}$, which is minimal along the first diagonal (i.e., when $\varepsilon_1=\varepsilon_2$) equaling $\Delta E=2|\mathcal{V}_{12}|$ then. Resulting from Eqs.~(\ref{def_EV1})-(\ref{def_v}), the spectral density $A_{11}(\varepsilon)$ in dot 1, respectively, $A_{22}(\varepsilon)$ in dot 2, is a linear combination of Dirac delta functions within a multiplicative factor $2\pi$, centered at the values of eigenenergies $E^+_\text{dots}$ and $E^-_\text{dots}$, with weighting factors equal to $|u|^2$ and $|v|^2$, respectively, $|v|^2$  and $|u|^2$  in dot 2. It has to be noted that as soon as the interdot coupling $\mathcal{V}_{12}$ becomes finite, a mixed spectral density $A_{12}(\varepsilon)$ arises, resulting from interdot transitions.

The charge stability diagram of the system can easily be derived from the latter results. At equilibrium the boundary lines between the domains of different occupations are obtained when any of the two levels of energies, $E^+_\text{dots}$ and $E^-_\text{dots}$ is aligned with the chemical potential of the leads $\mu_L=\mu_R=\mu$. The equations of the boundary lines  $\m{B}_+$ and $\m{B}_-$, hence given by $E^+_\text{dots}=\mu$ and $E^-_\text{dots}=\mu$, are $(\varepsilon_1-\mu)(\varepsilon_2-\mu)=\m{V}_{12}^2$. They correspond to two branches of an hyperbol in the plane $(\varepsilon_1, \varepsilon_2)$.The distance between the two branches $\m{B}_+$ and $\m{B}_-$  is minimal along the first diagonal taking the value of $2|\m{V}_{12}|$. The boundary lines are drawn in Figs.~\ref{figure_VP}(a), \ref{figure_VP}(c), and \ref{figure_VP}(e) at $\mu=0$ for different values of the interdot coupling $\m{V}_{12}$. One can check that the minimal distance between the two boundary lines  $\m{B}_-$ and  $\m{B}_+$ increases with increasing~$\m{V}_{12}$.

\subsection{ML-DQD}

 \begin{figure}
\hspace*{-0.5cm}\includegraphics[scale=.43]{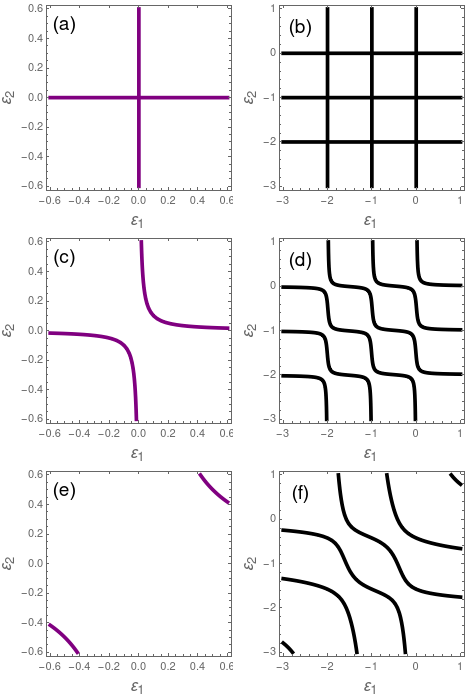}
\caption{Equienergetic curves of equations $E^\pm_\text{dots}=\mu_{L,R}$ (left column) and  $ E^\lambda_\text{dots}=\mu_{L,R}$ with $\lambda\in[1,6]$ (right column) at $\mu_{L,R}=0$ in the plane $(\varepsilon_1,\varepsilon_2)$ for (a)-(b) $\m{V}_{12}=0$, (c)-(d) $\m{V}_{12}=0.1$ and (e)-(f) $\m{V}_{12}=0.5$. The purple lines are obtained for a SL-DQD and the black lines for a ML-DQD with three levels of energy in each dot.}
\label{figure_VP}
\end{figure} 

For the three energy level ML-DQD considered in Sec.~\ref{section_ML_DQD_series}, the hamiltonian $\m{\widehat H}_\text{dots}$ writes as a $6\times 6$ matrix in the basis of the states $\{|1n\rangle,|2m\rangle\}$ with integer indices $n,m\in[0,2]$ in the two dots given by
\begin{eqnarray}
 \m{\widehat H}_\text{dots}=
\left(
\begin{array}{cccccc}
\varepsilon_{10} & 0 & 0 & \mathcal{V}_{12}^* & \mathcal{V}_{12}^* & \mathcal{V}_{12}^*\\
0 & \varepsilon_{11} & 0 & \mathcal{V}_{12}^* & \mathcal{V}_{12}^* & \mathcal{V}_{12}^*\\
0 & 0 &  \varepsilon_{12} & \mathcal{V}_{12}^* & \mathcal{V}_{12}^* & \mathcal{V}_{12}^*\\
\mathcal{V}_{21}^*&\mathcal{V}_{21}^*&\mathcal{V}_{21}^* & \varepsilon_{20} & 0 & 0 \\
\mathcal{V}_{21}^*&\mathcal{V}_{21}^*&\mathcal{V}_{21}^* & 0& \varepsilon_{21}  & 0 \\
\mathcal{V}_{21}^*&\mathcal{V}_{21}^*&\mathcal{V}_{21}^* & 0 & 0 &  \varepsilon_{22} \\
\end{array}
\right)
\end{eqnarray}
where $\varepsilon_{in}=\varepsilon_i+n\Delta\varepsilon_i$ and $\mathcal{V}_{12}$ is the interdot coupling of equal value regardless of the levels considered in each dot. $\m{\widehat H}_\text{dots}$ can be diagonalized leading to six eigenenergies $E_\text{dots}^\lambda$, with the integer index $\lambda\in[1,6]$, whose values can be numerically calculated. It is useful for the discussion in Sec.~\ref{section_ML_DQD_series} to determine the equienergetic curves at equilibrium: $E_\text{dots}^\lambda=\mu$. The results obtained numerically are displayed in Figs.~\ref{figure_VP}(b), \ref{figure_VP}(d), and \ref{figure_VP}(f) at $\mu=0$ for different values of the interdot coupling $\m{V}_{12}$. When $\m{V}_{12}=0$, the eigenenergies are simply equal to  $\varepsilon_1$, $\varepsilon_1+\Delta \varepsilon_1$, $\varepsilon_1+2\Delta \varepsilon_1$, $\varepsilon_2$, $\varepsilon_2+\Delta \varepsilon_2$ and $\varepsilon_2+2\Delta \varepsilon_2$, and the equienergetic curves at $ E_\text{dots}^\lambda =0$ are the three horizontal and three vertical lines of a square lattice as found in Fig.~\ref{figure_VP}(b). As soon as the interdot coupling $\m{V}_{12}$ gets finite, a level anticrossing effect takes place in the vicinity of the nodes of the square lattice, as shown in Figs.~\ref{figure_VP}(d) and \ref{figure_VP}(f). At $\m{V}_{12}= 0.1$, the equienergetic curves become sinuous, as a result of this level anticrossing effect. At $\m{V}_{12}=0.5$, the distance between two adjoining equienergetic curves increases, leaving room for wide interstitial areas in the direction parallel to the second diagonal.

These various elements brought by the above discussion are crucial to physically interpret the  results obtained  for the conductance, Seebeck coefficient, total dot occupancy, and zero-frequency charge susceptibility in the DQD system as discussed in Secs.~\ref{section_SL_DQD_series} and \ref{section_ML_DQD_series}.


\bibliographystyle{apsrev4-1}
\bibliography{references_8juin.bib}

\end{document}